\documentclass[twocolumn,trackchanges]{aastex631}

\usepackage{newtxtext,newtxmath}
\usepackage[T1]{fontenc}
\usepackage{ae,aecompl}
\usepackage{color}
\usepackage{comment}
\usepackage{ulem}

\usepackage{graphicx}	% Including figure files
\usepackage{amsmath}	% Advanced maths commands
\usepackage{amssymb}	% Extra maths symbols
\usepackage{mathrsfs}
\usepackage{enumerate}
\usepackage{hyperref}

\newcommand{\redrev}[1]{\textcolor{black}{ #1}}

\received{XXX XX, XXXX}
\revised{\today}
\accepted{XXXX}
\submitjournal{ApJ}

\shorttitle{Dark matter halo properties of the dwarf satellites}
\shortauthors{Hayashi et al.}
\begin{document}

\title{Dark matter halo properties of the Galactic dwarf satellites:\\ implication for chemo-dynamical evolution of the satellites and a challenge to $\Lambda$CDM}

\correspondingauthor{Kohei Hayashi}
\email{khayashi@sendai-nct.ac.jp}

\author[0000-0002-8758-8139]{Kohei Hayashi}
\affiliation{National Institute of Technology, Sendai College\\
48 Nodayama, Medeshima-Shiote, Natori, Miyagi 981-1239, Japan}
\affiliation{Astronomical Institute, Tohoku University \\
6-3 Aoba, Aramaki, Aoba-ku, Sendai, Miyagi 980-8578, Japan}
\affiliation{Institute for Cosmic Ray Research, The University of Tokyo \\
Chiba 277-8582, Japan}

\author[0000-0002-5661-033X]{Yutaka Hirai}
\altaffiliation{JSPS Research Fellow}
\affiliation{Astronomical Institute, Tohoku University \\
6-3 Aoba, Aramaki, Aoba-ku, Sendai, Miyagi 980-8578, Japan}
\affiliation{Department of Physics and Astronomy, University of Notre Dame \\
225 Nieuwland Science Hall, Notre Dame, IN 46556, USA}
\affiliation{Joint Institute for Nuclear Astrophysics, Center for the Evolution of the Elements (JINA-CEE), USA}

\author[0000-0002-9053-860X]{Masashi Chiba}
\affiliation{Astronomical Institute, Tohoku University \\
6-3 Aoba, Aramaki, Aoba-ku, Sendai, Miyagi 980-8578, Japan}

\author[0000-0002-5316-9171]{Tomoaki Ishiyama}
\affiliation{Digital Transformation Enhancement Council, Chiba University \\
1-33, Yayoi-cho, Inage-ku, Chiba 263-8522, Japan}

\begin{abstract}
Elucidating dark matter density profiles in the Galactic dwarf satellites is essential to understanding not only the quintessence of dark matter, but also the evolution of the satellites themselves.
In this work, we present the current constraints on dark matter densities in the Galactic ultra-faint dwarf (UFD) and diffuse galaxies.
Applying our constructed non-spherical mass models to the currently available kinematic data of the 25 UFDs and 2 diffuse satellites, we find that whereas most of the galaxies have huge uncertainties on the inferred dark matter density profiles, Eridanus~II, Segue~I, and Willman~1 favor cuspy central profiles even considering effects of a prior bias.
We compare our results with the simulated subhalos on the plane between the dark matter density at 150~pc and the pericenter distance.
We find that the most observed satellites and the simulated subhalos are similarly distributed on this plane, except for Antlia~2, Crater~2, and Tucana~3, which are less than one tenth of the density.
Despite considerable tidal effects, the subhalos detected by commonly-used subhalo finders have difficulty in explaining such a huge deviation.
We also estimate the dynamical mass-to-light ratios of the satellites and confirm the ratio is linked to stellar mass and metallicity.
Tucana~3 deviates largely from these relations, while it follows the mass-metallicity relation. This indicates that Tucana~3 has a cored dark matter halo, despite a significant uncertainty in its ratios.

%Using these scaling relations and the empirical tidal evolution models, we infer that the scaling laws can be sustained, and furthermore, a metallicity scatter at the faint-end of the mass-metallicity relation can be explained by tidal stripping of cuspy halos.

\end{abstract}

%\keywords{dark matter --- galaxies: dwarf --- galaxies: kinematics and dynamics --- galaxies: structure --- Local Group} 

%%%%%%%%%%%%%%%%%%%%%%%%%%%%%%%%%%%%%%%%%%%%%%%%%%%%%
%%%%% Introduction%%%%%%%%%%%%%%%%%%%%%%%%%%%%%%%%%%%
%%%%%%%%%%%%%%%%%%%%%%%%%%%%%%%%%%%%%%%%%%%%%%%%%%%%%
\section{Introduction} \label{sec:intro}
Dwarf spheroidal (dSph) galaxies associated with the Milky Way are valuable laboratories to shed light on the nature of dark matter because these galaxies are dark-matter dominated stellar systems, reaching dynamical mass-to-light ratios of up to $\sim10^3$~\citep[e.g.,][]{2012AJ....144....4M,2019ARA&A..57..375S}.
Since the dSphs are sufficiently nearby to resolve individual stars around the galaxies, it is possible to measure an accurate line-of-sight velocity and chemical composition for each star with higher-resolution spectroscopy~\citep[e.g.,][]{2007ApJ...670..313S,2009AJ....137.3100W}.
Therefore, utilizing these spectroscopic data, many early studies have estimated their dark matter distributions and dynamical masses through dynamical analysis~\citep[see][for a review]{2022NatAs.tmp..103B}.

%Owing to dedicated efforts across a wide range of observations such as photometry, spectroscopy, and astrometry, dwarf spheroidal galaxies~(dSphs) are the faint, old, metal-poor, chemically primitive, and dark-matter dominated stellar systems known. Therefore, these galaxies are ideal targets to shed light on the nature of dark matter and processes of galaxy formation at the early universe~[Refs.].

%Understanding the distributions of dark matter on galactic and sub-galactic scales is at the root of revealing the nature of dark matter. 
%The Galactic dSphs are valuable laboratories to understand the dark matter distribution on kpc scales.
%, because these satellites are dark-matter dominated system, reaching dynamical mass-to-light ratio of up to $\sim10^3$~[Refs.].

Recent dynamical studies for the luminous dSphs, so-called classical dSphs, have suggested that although there are still large uncertainties stemmed from insufficient data volume and dynamical modelings, these galaxies show a diversity of the inner dark matter densities~\citep[][]{2019MNRAS.484.1401R,2019MNRAS.490..231K,2020ApJ...904...45H}.
Furthermore, this diversity could be explained by the $\Lambda$-cold dark matter~($\Lambda$CDM) framework combined with baryonic feedback and star formation burst predicted by recent $N$-body and hydrodynamical simulations~\citep[e.g.,][]{2016MNRAS.456.3542T,2020MNRAS.497.2393L}.
Thus, the core-cusp problem, which is one of the controversial issues in $\Lambda$CDM models~\citep[][]{2017ARA&A..55..343B}, can be ameliorated by a baryonic feedback mechanism.
Meanwhile, we should bear in mind that other dark matter candidates are still attractive models.
For instance, self-interacting dark matter (SIDM) models could also reproduce the diversity by considering a gravothermal instability induced by tidal stripping~\citep[e.g.,][]{PhysRevD.101.063009,2021MNRAS.503..920C}.
Consequently, from recent observational and computational studies, such luminous dSphs could not hold the pristine dark-matter distributions.

In the ultra-faint dwarf (UFD) galaxies, which are much fainter and less-massive than the classical ones, the impact of baryonic effects on their central dark matter densities can be negligible.
Thus the UFDs are promising targets to set constraints on dark matter models through the dark matter distributions.
Although the number of observable individual stars are limited because most of their member stars are fainter than the magnitude limits of the current spectroscopy, several studies made an attempt to infer their enclosed masses of dark matter halos and set a constraint on critical parameters for several dark matter models~\citep[e.g.,][]{2018MNRAS.481.5073E,2021ApJ...912L...3H,2021PhRvD.103b3017H}.
However, the central dark matter density profiles in the UFDs are still unknown.

%In light of galaxy formation and evolution, the UFDs as well as the luminous dSphs have drawn attention as building blocks of bright host galaxies, thereby enabling us to unearth a fossil signature in the evolution of the Galaxy and the Local Group~[Refs.].
%The UFDs, in particular, provide unique windows into the formation and evolution of the first galaxies before the era of reionization.
%This is because from a wide range of observations such as photometry, spectroscopy, and astrometry, these galaxies can be the faintest, oldest, most metal-poor and chemically primitive known~[Refs.].

%From high-resolution galaxy simulations, the formation of the first galaxies depends largely on the gas temperatures which can be cool enough for star formation to begin~[Refs.].
%Such threshold temperatures correspond generally either to the dark matter halos of $10^6-10^8M_\odot$, which cool through molecular hydrogen at $z\sim20$, or in atomic cooling halos of heavier than $10^8M_\odot$, which cool initially through atomic hydrogen lines and collapse later at $z\sim10$~[Refs.].
%UFDs may form in the dark matter halos which meet either halo mass threshold. 
%However, due to the high resolution and wide dynamic range required, simulations for such faint satellites have a difficult computational issue.
%There are dedicated efforts to address the issue~[Refs.], but much more computational power and higher resolution are still needed to study formation of UFDs in detail.

The currently discovered UFDs have been orbiting around the Milky Way and thus are thought to have experienced tidal stripping and/or disruption.
Thanks to the Gaia astrometry~\citep{2018A&A...616A..12G,2021A&A...649A...1G}, the bulk proper motions of all UFDs can be determined accurately, and then their orbits are computed with their known radial velocities~\citep[][]{2020RNAAS...4..229M,2021ApJ...916....8L,2022A&A...657A..54B}.
For a given model of the gravitational potential of the Milky Way, the pericenter distance is generally determined within $10-20$\% for the satellites within 100~kpc from the Sun, and thereby offering the opportunity to study the evolution of the Galactic satellites~\citep[e.g.,][]{2020ApJ...905..109M,2020arXiv201109482G,2021MNRAS.506..531D}.
From theoretical studies, while tidal stripping removal of dark matter in the satellites is inevitable, an appreciable fraction of their stars can be lost only if they approach the center of the Milky Way very closely~\citep{2008ApJ...673..226P}.
In fact, several galaxies show signs of the tidal effects from morphology, possible velocity gradients and so on~\citep[e.g.,][]{2015ARA&A..53..631F,2019ARA&A..57..375S}.

In terms of tidal effects, Antlia~2 and Crater~2 have drawn attention as unusually diffuse satellite galaxies in the Milky Way~\citep{2016MNRAS.459.2370T,2019MNRAS.488.2743T}.
More surprisingly, these galaxies have exceptionally low velocity dispersion, even though they have large half-light radii~\citep[][]{2017ApJ...839...20C,2021ApJ...921...32J}.
A central question is whether such diffuse galaxies can be explained by tidal forces in the context of $\Lambda$CDM paradigm.
Several studies based on simulations indicate that $\Lambda$CDM models can reproduce the stellar properties of the diffuse galaxies with strong tidal stripping, especially if the dark matter halo has cored central density induced by stellar feedback and/or the halo has extremely low concentration~\citep[e.g.,][]{2015MNRAS.449L..46E,2019ApJ...886L...3R,2019MNRAS.488.2743T,2020ApJ...892..137S,2022NatAs...6..496M}.
Therefore, elucidating their central densities affords the key to understanding the formation history of the Galactic diffuse galaxies. 

Given the current situations detailed above, in order to understand the formation of the Galactic satellites, it is essential to consider not only stellar properties such as luminosity and metallicity but also dynamical ones including dark matter distributions. 
For instance, the mass-metallicity relation~\citep[][hereafter MZR]{2013ApJ...779..102K}, which is a tight correlation between stellar mass and mean stellar metallicity, can provide us hints for the formation of the satellites.
This relation argues that tidal stripping of their stellar components could not be effective, because this effect reduces the luminosity of a galaxy without noticeably changing its metallicity in the absence of strong metallicity gradients.
On the other hand, a large scatter of metallicity at the faint-end of this relation surely exists~\citep{2019ARA&A..57..375S}. 
The possible reason for this scatter is tidal stripping, but it is not conclusive because of a lack of knowledge of dark matter content and extent in the UFDs.
Thus, in order to get insight into the roots of this scatter, knowledge of the dark matter halo properties of the UFDs should be essential.

Motivated by the aforementioned standing situation, we analyze the currently available kinematics for 25 UFDs and 2 diffuse satellites to estimate dark matter density profiles and dynamical masses through non-spherical dynamical models based on axisymmetric Jeans equations constructed by our group~\citep{2012ApJ...755..145H,2015ApJ...810...22H,2020ApJ...904...45H}.
The main reason why we employ the non-spherical mass models is that such models can treat two-dimensional distributions of line-of-sight velocity dispersions and thereby mitigating a strong degeneracy between dark matter density and stellar anisotropy in spherical models~\citep{1982MNRAS.200..361B}.

This paper is organized as follows.
In Section~\ref{sec:model_data}, we briefly explain our dynamical models, and then we describe the photometric and spectroscopic data for our sample dwarf galaxies and our fitting procedure.
In Section~\ref{sec:results}, we present the results of the fitting analysis and the estimated dark matter density profiles based on the fitting results.
We also show the relation between inner dark matter density slope and stellar-to-halo mass ratio predicted by recent galaxy simulations.
In Section~\ref{sec:Discussion}, using our estimated dark matter halo properties, such as the dark matter central densities and dynamical mass-to-light ratios, we discuss the tidal evolution of the Galactic satellites and the implication for a challenge to $\Lambda$CDM paradigm.
Finally, the summary and conclusion are presented in Section~\ref{sec:Conclusion}.

%%%%%%%%%%%%%%%%%%%%%%%%%%%%%%%%%%%%%%%%%%%%%%%%%%%%%
%%%%% Models and data %%%%%%%%%%%%%%%%%%%%%%%%%%%%%%%
%%%%%%%%%%%%%%%%%%%%%%%%%%%%%%%%%%%%%%%%%%%%%%%%%%%%%
\section{Models and data} \label{sec:model_data}
In this section, we briefly introduce our dynamical mass models.
In this work, to estimate the dark matter density profiles and dynamical masses in the Galactic satellites, we use the same dynamical analysis as~\citet[][]{2020ApJ...904...45H}. 
Thus, the further details about this analysis are found in that article.

%%% Table 1 %%%
\begin{table*}
	\centering
	\caption{The observational data for the ultra-faint dwarf galaxies. \redrev{The bold name of galaxies are those which are shown in Figure~\ref{fig:DMprofiles}.}}
	\label{table1}
	\begin{tabular}{lrcccccl} % four columns, alignment for each
		\hline\hline
Object 
& $N_{\rm sample}$ 
& $\log_{10}(L_V/L_\odot)$        
& $D_{\odot}$ 
& $b_{\ast}$ 
& $q^{\prime}$  
& $\langle \mathrm{[Fe/H]}\rangle$ 
%& $\sigma_\mathrm{[Fe/H]}$
&  Reference\\
&                  
&
& [kpc]       
& [pc]       
& (axial ratio) 
& [dex]  
%& [dex]
& \\
\hline
{\bf Antlia~2} %(Ant2)
& 283    
& $5.88\pm0.10$ 
& $129\pm7$  
& $2867\pm312$  
& $0.62\pm0.08$ 
& $-1.77\pm0.08$ 
%& $0.66\pm0.03$
& (1), (9)\\
{\bf Bo$\ddot{\mathrm{o}}$tes I} %(Boo1)
& 66    
& $4.33\pm0.10$ 
& $ 66\pm2$  
& $191\pm8$  
& $0.70\pm0.03$ 
& $-2.34\pm0.05$ 
%& $0.37\pm0.08$
& (2), (3), (4)\\
{\bf Canes Venatici I} %(CVn1)
& 214   
& $5.45\pm0.02$ 
& $218\pm2$  
& $452\pm13$  
& $0.61\pm0.03$ 
& $-1.91\pm0.01$   
%& $0.44$
&  (2), (5), (6)\\
Canes Venatici II %(CVn2)      
& 25    
& $4.00\pm0.13$ 
& $160\pm2$  
& $71\pm11$  
& $0.60\pm0.13$ 
& $-2.12\pm0.05$
%& $0.59$
&  (2), (5), (7)\\
{\bf Coma  Berenices} %(CB)     
& 59    
& $3.68\pm0.10$ 
& $44\pm2$  
& $72\pm4$  
& $0.62\pm0.05$ 
& $-2.25\pm0.05$
%& $0.43$
&  (2), (5), (8)\\
{\bf Crater 2}  %(Cra2)
& 141    
& $5.21\pm0.10$ 
& $118\pm1$  
& $1066\pm84$  
& $0.88\pm0.02$ 
& $-2.10\pm0.08$  
%& $0.34\pm0.03$
& (9), (10)\\
Draco 2 %(Dra2)        
& 9    
& $3.10\pm0.40$ 
& $20\pm3$  
& $19^{+8}_{-6}$  
& $0.76\pm0.25$ 
& $-2.70\pm0.1$   
%& $<0.24$
&  (11), (12)\\
{\bf Eridanus II} %(Eri2)
& 92    
& $4.82\pm0.04$  
& $380\pm2$  
& $196\pm19$  
& $0.65\pm0.06$ 
& $-2.38\pm0.13$   
%& $0.47^{+0.12}_{-0.09}$
&  (2), (13), (14)\\
Grus 1 %(Gru1)   
& \redrev{8}   
& $3.32\pm0.24$  
& $120\pm11$  
& $28\pm23$  
& $0.55\pm0.3$ 
& \redrev{$-2.62\pm0.1$}%$-1.88^{+0.09}_{-0.03}$   
%& $0.00^{+0.90}_{-0.00}$
&  (2), \redrev{(15)}\\
Grus 2 %(Gru2)        
& \redrev{19}   
& $3.33\pm0.04$   
& $55\pm2$  
& $94\pm9$  
& $1.00^{+0.0}_{-0.21}$ 
& $-2.51\pm0.11$   
%& $0.00^{+0.45}_{-0.00}$
&  (16), (17), (18)\\
Hercules %(Her)      
& 18   
& $4.27\pm0.07$   
& $132\pm12$  
& $216\pm17$  
& $0.31\pm0.03$ 
& $-2.39\pm0.04$   
%& $0.51$
&  (2), (5), (19)\\
Horologium I  %(Hor1)       
& 5   
& $3.35\pm0.22$   
& $79\pm2$  
& $37\pm7$  
& $0.73\pm0.13$ 
& $-2.76\pm0.1$   
%& $0.17^{+0.20}_{-0.03}$
&  (2), (20)\\
Hydra II  %(Hyd2)       
& 13   
& $3.77\pm0.15$   
& $134\pm10$  
& $59\pm11$  
& $0.76\pm0.16$ 
& $-2.02\pm0.08$   
%& $0.40^{+0.48}_{-0.26}$
&  (2), (21), (22)\\
Leo IV        
& 18   
& $3.93\pm0.10$   
& $154\pm5$  
& $114\pm13$  
& $0.83\pm0.09$ 
& $-2.47\pm0.14$   
%& $0.42^{+0.12}_{-0.09}$
&  (2), (3), (23)\\
Leo V        
& 7   
& $3.69\pm0.14$   
& $169\pm4$  
& $49\pm16$  
& $0.57\pm0.22$ 
& $-2.28\pm0.16$   
%& $0.34^{+0.17}_{-0.10}$
&  (2), (3), (23) \\
Leo T       
& 19   
& $4.97\pm0.06$   
& $417\pm19$  
& $153\pm16$  
& $0.77\pm0.09$ 
& $-1.74\pm0.04$   
%& $0.54$
&  (2), (5), (24)\\
Pisces II %(Pis2)        
& 7   
& $3.62\pm0.15$   
& $182\pm2$  
& $59\pm9$  
& $0.66\pm0.10$ 
& $-2.45\pm0.07$   
%& $0.48^{+0.70}_{-0.29}$
&  (2), (21), (25) \\
Reticulum II %(Ret2)        
& 25   
& $3.48\pm0.15$   
& $32\pm2$  
& $48\pm2$  
& $0.42\pm0.02$ 
& $-2.46\pm0.1$   
%& $0.29^{+0.13}_{-0.05}$
&  (2), (20), (26)\\
{\bf Segue 1} %(Seg1)
& 71   
& $2.45\pm0.29$   
& $23\pm2$  
& $24\pm3$  
& $0.67\pm0.1$ 
& $-2.71^{+0.45}_{-0.39}$   
%& $0.95^{+0.42}_{-0.26}$
&  (2), (19), (27)\\
Segue 2 %(Seg2)    
& 26   
& $2.77\pm0.35$   
& $35\pm2$  
& $38\pm3$  
& $0.78\pm0.07$ 
& $-2.22\pm0.13$   
%& $0.43$
&  (2), (28), (29) \\
Triangulum II %(Tri2)      
& 13   
& $2.65\pm0.20$  
& $30\pm2$  
& $17\pm4$  
& $0.54\pm0.16$ 
& $-2.24\pm0.05$   
%& $0.53^{+0.38}_{-0.12}$
&  (2), (30), (31)\\
Tucana 2 %(Tuc2)       
& \redrev{19} 
& $3.45\pm0.63$ 
& $57\pm5$ 
& $165^{+28}_{-19}$ 
& $0.61\pm0.15$ 
& $-2.23^{+0.18}_{-0.12}$  
%& $0.29^{+0.15}_{-0.12}$
&   (26), \redrev{(35)}\\
Tucana 3 %(Tuc3)       
& 26   
& $2.70\pm0.10$  
& $25\pm2$ 
& $44\pm6$ 
& $0.80\pm0.1$ 
& $-2.42\pm0.08$  
%& $<0.19$
&  (16), (32)\\
Tucana 4 %(Tuc4)       
& 11
& $3.14\pm0.07$
& $48\pm4$ 
& $127\pm24$
& $0.40\pm0.1$
& $-2.49\pm0.16$
%& $0.00^{+0.64}_{-0.00}$
&  (16), (18)\\
Ursa Major I %(Uma1)       
& 39 
& $3.98\pm0.15$
& $97\pm4$
& $234\pm10$ 
& $0.41\pm0.03$ 
& $-2.10\pm0.03$ 
%& $0.65$
&  (2), (5), (33)\\
Ursa Major II %(Uma2)      
& 20 
& $3.63\pm0.10$ 
& $32\pm4$
& $128\pm5$  
& $0.44\pm0.03$
& $-2.18\pm0.03$ 
%& $0.66$
& (2), (5), (33)\\
{\bf Willman 1} %(Will1)
& 40
& $2.94\pm0.30$
& $38\pm7$
& $28\pm2$
& $0.53\pm0.06$
& $-2.19\pm0.08$
%& $0.56$
&  (2), (34)\\
	\hline
	\end{tabular}
\begin{flushleft}
The table lists the number of member stars with velocity measurements available from the kinematic analysis~($N_\mathrm{sample}$), V-band absolute magnitudes~($L_V$), distances from the Sun~($D_\odot$), projected half-light radii~($b_\ast$), projected stellar axial ratios~($q^\prime$), mean metallicities~($\langle\mathrm{[Fe/H]}\rangle$), and references. Here, a projected stellar axial ratio $q^\prime$ is related to the intrinsic one~$q$ via the inclination angle~$i$: $q^{\prime2}=\cos^2i+q^2\sin^2i$. This equation can be rewritten as $q=\sqrt{q^{\prime2}-\cos^2i}/\sin i$ \citep{1926ApJ....64..321H}. Thus, the range of inclination angle is confined by $0\leq\cos^2i\leq q^{\prime2}$.

References: (1)~\citet{2019MNRAS.488.2743T};
(2)~\citet{2018ApJ...860...66M};
(3)~\citet{2021arXiv210100013J};
(4)~\citet{2006ApJ...653L.109D};
(5)~\citet{2013ApJ...779..102K};
(6)~\citet{2008ApJ...674L..81K};
(7)~\citet{2008ApJ...675L..73G};
(8)~\citet{2007ApJ...654..897B};
(9)~\citet{2021ApJ...921...32J};
(10)~\citet{2016MNRAS.459.2370T};
(11)~\citet{2015ApJ...813...44L};
(12)~\citet{2018MNRAS.480.2609L};
(13)~\citet{2021A&A...651A..80Z};
(14)~\citet{2017ApJ...838....8L};
(15)~\citet{2022arXiv220604580C};
(16)~\citet{2015ApJ...813..109D};
(17)~\citet{2019MNRAS.490.2183M};
(18)~\citet{2020ApJ...892..137S};
(19)~\citet{2008ApJ...684.1075M};
(20)~\citet{2015ApJ...811...62K};
(21)~\citet{2015ApJ...810...56K};
(22)~\citet{2015ApJ...804L...5M};
(23)~\citet{2010ApJ...710.1664D};
(24)~\citet{2008ApJ...680.1112D};
(25)~\citet{2010ApJ...712L.103B};
(26)~\citet{2015ApJ...805..130K};
(27)~\citet{2011ApJ...733...46S};
(28)~\citet{2009MNRAS.397.1748B};
(29)~\citet{2013ApJ...770...16K};
(30)~\citet{2015ApJ...802L..18L};
(31)~\citet{2017ApJ...838...83K};
(32)~\citet{2017ApJ...838...11S};
(33)~\citet{2007ApJ...670..313S};
(34)~\citet{2011AJ....142..128W};
(35)~\citet{2022arXiv220501740C}
\end{flushleft}
\end{table*}

\subsection{Non-spherical mass models}
Assuming that a galaxy is an axisymmetric and dark matter dominated steady-state system, the stellar motion of such a system can be described by the axisymmetric Jeans equations, which are taken moments of the steady-state collisionless Boltzmann equation~\citep{2008gady.book.....B}.
\redrev{We assume that the cross terms of velocity moments such as $\overline{u_Ru_z}$ vanish and the velocity ellipsoid constituted by $(\overline{u^2_R},\overline{u^2_{\phi}},\overline{u^2_z})$ is aligned with the cylindrical coordinate.
We also assume that the density of tracer stars has the same orientation and symmetry as that of a dark halo and that
$\beta_z=1-\overline{u^2_z}/\overline{u^2_R}$ is a constant velocity anisotropy parameter introduced by~\citet{2008MNRAS.390...71C}. 
Owing to large mass-to-light ratios of the dSphs, we assume the stellar mass to be negligible. 
Under these assumptions,
the Jeans equations are expressed as}
\begin{eqnarray}
\overline{u^2_z} &=&  \frac{1}{\nu(R,z)}\int^{\infty}_z \nu\frac{\partial \Phi_\mathrm{DM}}{\partial z}dz,
\label{AGEb03}\\
\overline{u^2_{\phi}} &=& \frac{1}{1-\beta_z} \Biggl[ \overline{u^2_z} + \frac{R}{\nu}\frac{\partial(\nu\overline{u^2_z})}{\partial R} \Biggr] +
R \frac{\partial \Phi_\mathrm{DM}}{\partial R},
\label{AGEb04}
\end{eqnarray}
where $\nu$ is the three-dimensional stellar density and $\Phi_\mathrm{DM}$ is the dark matter gravitational potential.
%\footnote{Nevertheless, this assumption is roughly in good agreement with dark matter simulations reported by~\citet{2014MNRAS.439.2863V} who have shown that simulated subhalos have an almost constant $\beta_z$ or a weak trend as a function of radius along each axial direction.}.
In principle, these second velocity moments are defined as $\overline{u^2}= \sigma^2+\overline{u}^2$, where $\sigma$ and $\overline{u}$ are dispersion and streaming motions of stars, respectively.
Since the UFDs are largely dispersion-supported stellar systems~\citep[e.g.,][]{2017MNRAS.465.2420W}, the latter streaming motions can be negligible.
To compare observed line-of-sight velocity dispersions with the models, we project the intrinsic velocity dispersions from the equations~(\ref{AGEb03}) and (\ref{AGEb04}) into the line-of-sight direction followed by previous works~\citep{1997MNRAS.287...35R,2006MNRAS.371.1269T,2012ApJ...755..145H}.

For the stellar density profile, we adopt the Plummer profile~\citep{1911MNRAS..71..460P} generalized to an axisymmetric shape, $\nu\propto(3/4\pi b^3_\ast)(1+r_\ast^2/b^2_\ast)^{-5/2}$, where $r_\ast^2=R^2+z^2/q^2$ and $b_\ast$ is a half-light radius along the major axis. 
$q$ is an intrinsic axial ratio of the stellar distribution assumed to be independent of position. The surface density profile can be derived from $\nu$ via an Abel transform and the inclination angle, $i$, i.e., the angle between the symmetry axis of the dwarf and the line of sight, which is a free parameter. When a galaxy is edge-on (face-on), $i=90^\circ$ ($i=0^\circ$), respectively.

Meanwhile, to determine the dark matter gravitational potential, we assume a generalized Hernquist profile~\citep{1990ApJ...356..359H,1996MNRAS.278..488Z} but non-spherical shape,
\begin{eqnarray}
&& \rho_{\rm DM}(R,z) = \rho_0 \Bigl(\frac{r}{b_{\rm halo}} \Bigr)^{-\gamma}\Bigl[1+\Bigl(\frac{r}{b_{\rm halo}} \Bigr)^{\alpha}\Bigr]^{-\frac{\beta-\gamma}{\alpha}},
 \label{DMH} \\
&& r^2=R^2+z^2/Q^2,
\label{DMH2}
\end{eqnarray}
where $\rho_0$ and $b_{\rm halo}$ are the scale density and radius, respectively; $\alpha$ is
the sharpness parameter of the transition from the inner slope $\gamma$ to the outer slope $\beta$; and $Q$ is a constant axial ratio of a dark matter halo.
These $(Q, \rho_0, b_{\rm halo}, \alpha, \beta, \gamma)$ are the free parameters in our models.

\subsection{Data and Analysis}
\label{sec:dataanalysis}
In this work, we scrutinize dark matter halos for 25 UFDs~(Bo\"otes~I, Canes Venatici~I, Canes Venatici~II, Coma Berenices, Draco~2, Eridanus~II, Grus~1, Grus~2, Hercules, Horologium~I, Hydra~II, Leo~IV, Leo~V, Leo~T\footnote{This galaxy is actually a dwarf irregular galaxy, but we include it in this work because this galaxy has drawn attention as an ideal target to set constraints on dark matter models, especially for primordial black holes~\citep[e.g.,][]{2013ApJ...777..119F,2021PhRvD.103l3028W,2021ApJ...908L..23L}.}, Pisces~II, Reticulum~II, Segue~1, Segue~2, Triangulum~II, Tucana~2, Tucana~3, Tucana~4, Ursa~Major~I, Ursa~Major~II, Willman~1) and two diffuse galaxies~(Antlia~2 and Crater~2) in the Milky Way. The basic stellar properties are shown in Table~\ref{table1}.

The stellar structural parameters~($D_\odot$, $b_\ast$ and $q^\prime$ with the Plummer profile) of the galaxies are adapted from the original observation articles.
For the stellar kinematics of their member stars, we use the currently available data taken from each spectroscopic observation article.
For the membership selections for each galaxy, we also adopt the methods in the spectroscopic observation articles. 
For the influence of unresolved binary stars on stellar kinematics, several articles indicated that multi-epoch observations can exclude binary candidates from stellar spectroscopic data~\citep[e.g.,][]{2007ApJ...670..313S,2017ApJ...838...83K,2021arXiv210100013J}. 
Therefore, we suppose that the effect of binaries can be negligible.

To perform fitting analysis compering our mass models with the kinematic data of our sample galaxies, we employ the likelihood function assuming that the line-of-sight velocity distribution of each star is a Gaussian shape and centered on the mean velocity of the galaxy~$\langle u \rangle$\footnote{In this work, we only consider a dispersion of the line-of-sight velocity distribution, hence we assume that this distribution is a Gaussian. However, we should bear in mind that there is still degeneracy between mass distribution and velocity anisotropy under this assumption. 
To mitigate the degeneracy, information about the shape of velocity distribution function (i.e., kurtosis) should be important.
Several papers~\citep[e.g.,][]{1984ApJ...286...27R, 2002MNRAS.333..697L, 2013MNRAS.429.3079M,2017MNRAS.471.4541R} constructed spherical mass models with taking into account higher-order moments. However, the current available data is possibly too sparse to break the degeneracy because the shape of velocity distribution is sensitive to the data volume and quality.}:
\begin{equation}
-2\ln(\mathcal{L}) = \sum_{i}\left[\frac{(u_i-\langle u \rangle)^2}{\sigma^2_i} + \ln(2\pi\sigma^2_i)\right]\,,
\label{eq:likeli_1}
\end{equation}
where $u_i$ is the line-of-sight velocity of the $i^{\mathrm th}$ star in the kinematic sample. 
The mean velocity of the galaxy~$\langle u \rangle$ is a nuisance parameter.
The dispersion $\sigma^2_i$ can be written by $\sigma^2_i=\delta^2_{v,i}+\sigma^2_{\mathrm{los},i}$, where $\delta_{v,i}$ the measurement error of the velocity, and $\sigma^2_{\mathrm{los},i}$ is the 
theoretical line-of-sight velocity dispersion at the sky plane position $(x_i,y_i)$, which is computed by model parameters and the Jeans equations.

\begin{figure*}
\begin{center}
 \includegraphics[width=\textwidth]{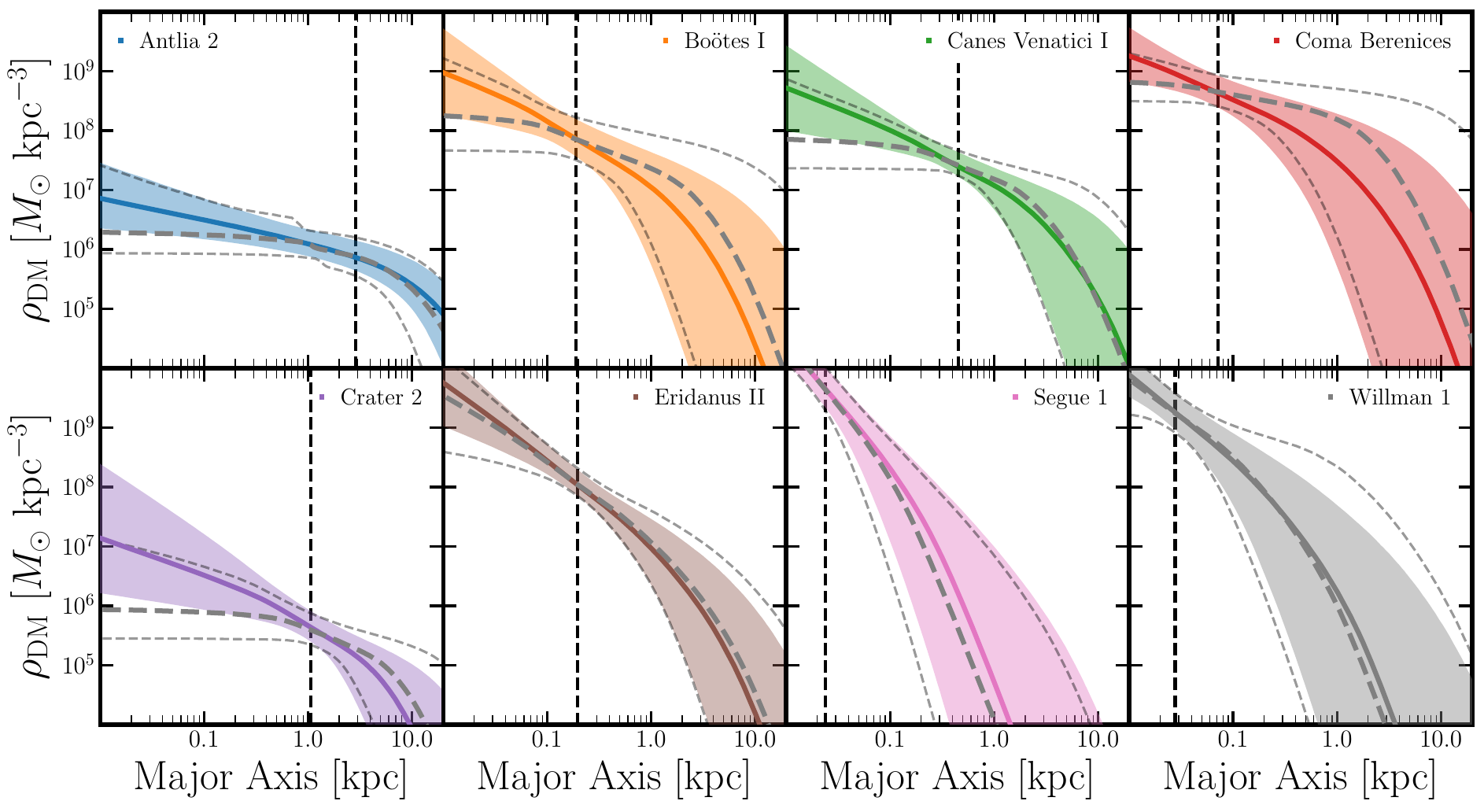}
\end{center}
\caption{Inferred dark matter density profiles along the major axes of the eight galaxies having the largest numbers of kinematic sample.
The solid line in each panel denotes the median value, and the shaded region depicts the 68\% confidence interval.
On the other hand, the gray thick and thin dashed lines show the median and the the 68\% confidence intervals for the case of a wider prior range~($-2.0\leq\gamma^\prime\leq2.0$, but if $\gamma^\prime<0 \rightarrow\gamma=0$, see the text in details).
The vertical dashed line in each panel corresponds to the half-light radius of each galaxy.}
\label{fig:DMprofiles}
\end{figure*}

Using the likelihood function~$\mathcal{L}$, we fit the 8 parameters~$(Q,b_\mathrm{halo},\rho_0,\beta_z,\alpha,\beta,\gamma,i)$ based on Bayesian statistics.
We estimate the posterior distributions of these parameters via Markov chain Monte Carlo by the Metropolis–Hastings algorithm~\citep{1953JChPh..21.1087M,10.1093/biomet/57.1.97}\footnote{We use our own custom-made code for our analysis. To assess the convergence of the Markov Chain Monte Carlo (MCMC) algorithm, we employ a visual inspection method, among several available methods. In our MCMC analysis, we utilize 20 chains, with each chain containing a sample size of 50,000 after discarding the burn-in phase. Specifically, we discard 5,000 samples from each chain as burn-in. Consequently, we have a total of one million samples, excluding the burn-in phase.}.
We adopt flat or log-flat priors over the following ranges: 
$0.1\leq Q\leq 2$, 
$0\leq \log_{10} [b_\mathrm{halo}/\mathrm{pc}]\leq 5$,
$-5\leq \log_{10} [\rho_0/(M_\odot/\mathrm{pc}^{-3})]\leq 5$,
$-1\leq-\log_{10}[1-\beta_z]<1$,
$0.5\leq\alpha\leq3$,
$3\leq\beta\leq10$,
$0\leq\gamma\leq2$, and
$\cos^{-1}(q^\prime)<i/\mathrm{deg}\leq90$.

%%%%%%%%%%%%%%%%%%%%%%%%%%%%%%%%%%%%%%%%%%%%%%%%%%%%%
%%%%% Results %%%%%%%%%%%%%%%%%%%%%%%%%%%%%%%%%%%%%%%
%%%%%%%%%%%%%%%%%%%%%%%%%%%%%%%%%%%%%%%%%%%%%%%%%%%%%

%\begin{figure*}
%\begin{center}
% \includegraphics[width=\textwidth]{./8UFDs_new_DMprofile_new.pdf}
%\end{center}
%\caption{Inferred dark matter density profiles along the major axes of the eight galaxies having the largest numbers of kinematic sample.
%The solid line in each panel denotes the median value, and the dark and light shaded regions depict the 68\% and 95\% confidence intervals.
%The vertical dashed line in each panel corresponds to the half-light radius of each galaxy.}
%\label{fig:DMprofiles}
%\end{figure*}

\section{Results} \label{sec:results}
In this section, we present the results from the MCMC fitting analysis described above, especially for their dark matter density profiles.
Furthermore, using the estimated dark matter density profiles, we calculate the dynamical mass-to-light ratios within their half-light radius and the dark matter densities at 150~pc.

\subsection{Parameter estimation}
We sampled the parameter spaces of our non-spherical mass models using the parametrizations described above.
The best-fitting parameters for each galaxy are shown in Table~\ref{table2} in Appendix~\ref{sec:appA}, including the 68\% confidence intervals.
As a by-product of the fitting results, we also show the dynamical mass-to-light ratio, $M_\mathrm{dyn}/L$, within half-light radius and the dark matter density at 150~pc, $\rho_\mathrm{DM}(150~\mathrm{pc})$, computed from the estimated dark matter halo parameters. 
Moreover, we show the Bayes factors, which are the ratio of the mean posterior distribution of the mass models with the spherical dark matter halo $(Q=1)$ and with all free parameters (subscript ``1''), and the ratio of the posterior distribution of the mass models with non-spherical dark matter halo~$(Q=q^{\prime})$ and with all free parameters (subscript ``2'').
In each model, we keep the stellar distributions have an axisymmetric shape, that is, stellar axial ratios are not unity.
The Bayes factors and Akaike Information Criterion show that there is no statistical evidence for the need to an additional free parameter for the halo axis ratio $Q$, compared to either spherical ($Q=1$) or equal to that of the stars ($Q=q^\prime$) for small sample-sized dSphs.

From the values of their Bayes factors, it is difficult to determine which spherical or non-spherical dark matter halo is better based on the currently available data.

Figures~\ref{PDFs1} and \ref{PDFs2} display the posteriors for all parameters for Antlia~2, Bo\"otes~I, Canes Venatici~I, Coma~Berenices, Crater~2, Eridanus~II, Segue~1, and Willman~1 as the representative satellites, which have the largest samples among our studied dwarfs. 
The results for other satellites are similar to these or are wider distributed in each parameter range.
The contours in these figures show 68\%, 95\%, and 97\% confidence intervals. The vertical lines in each histogram also depict the median and 68\% confidence upper and lower limits.
According to the maps, the parameters $Q, \alpha, \beta$, and $i$ are widely distributed in these parameter ranges, due to an insufficient number of spectroscopic data. 
It is confirmed that $b_\mathrm{halo}$-$\rho_0$ and $Q$-$\beta_z$ are clearly degenerate.
These results have already known and been discussed in several previous articles~\citep[e.g.,][]{2008gady.book.....B,2008MNRAS.390...71C,2015ApJ...801...74G,2015ApJ...810...22H,2020ApJ...904...45H}.
The inner slope parameter of the dark matter density profile $\gamma$ also shows a wide spread within its prior range.
Furthermore, the inner slope appears to be uncorrelated with the other two DM density profile parameters $\alpha$ and $\beta$, hence there is no need to fix $\alpha$ and/or $\beta$.
Nonetheless, several UFDs~(Eridanus II, Segue~1, and Willman~1) favor cuspy dark matter density profiles.
We discuss this further in the next section.

\subsection{Dark matter density profile}
\subsubsection{Profile Recovery}
Marginalizing the posteriors of the parameters, we recover the dark matter density profiles of the dwarf galaxies.
Figure~\ref{fig:DMprofiles} shows the inferred dark matter density profiles of the representative galaxies.
The solid lines show the median, and the shaded region marks the 68\% interval.
The vertical dashed lines are the projected half-light radii of each galaxy.

Firstly, Segue~1 and Willman~1 favor a sharply cuspy inner slope of their dark matter density profiles~($\gamma\sim1.6$ for Segue~1 and $\sim1.2$ for Willman~1), even considering 95\% confidence intervals of the profile.
Eridanus~II also prefers a cuspy dark matter density profile~($\gamma\sim1.2$), even though the uncertainty for its $\gamma$ is larger than for Segue~1 and Willman~1.
Intriguingly, the slope of Segue~1 is much steeper than an NFW cusp~($\gamma=1$) but is roughly consistent with ones predicted by the other simulation works~\citep[e.g.,][]{1997ApJ...477L...9F,1999MNRAS.310.1147M,2008MNRAS.391.1685S,2020MNRAS.492.3662I}, while those of Eridanus~II and Willman~1 are consistent with the NFW ones.
\redrev{Figure~\ref{fig:DMprofiles} indicates that the uncertainties on the mass density profile are narrowest at near the half-light radius, consistent with \citet{2010MNRAS.406.1220W}.}
Moreover, several studies have reported that these UFDs, especially Segue~1 and Willman~1 can provide the strongest constraint on fuzzy dark matter~(FDM) and self-interacting (SIDM) models which can create a naturally cored dark matter density profile without relying on any baryonic physics~\citep[e.g.,][]{2021ApJ...912L...3H,2021PhRvD.103b3017H}.
This implies that Segue~1 and Willman~1 could have cuspy dark matter halos.

Secondly, the other well-sampled UFDs tend to be somewhat cuspy profiles, but there are huge uncertainties in the inferred density slopes.
In fact, they permit $\gamma\sim0$, that is, a cored dark matter density within 95\% confidence.
Therefore, it is impossible to make a final conclusion whether these remaining poorly-sampled galaxies have cored or cuspy dark matter halos.
The other galaxies, which are not shown here, also have huge uncertainties on their inner slopes of dark matter density profiles from the 8th column in Table~\ref{table2}, which shows the constraints on $\gamma$. 

Thirdly, dark matter densities of the diffuse galaxies~(Antlia~2 and Crater~2) are around one order of magnitude less dense than those of the other dwarf satellites.
In particular, Antlia~2 favors having the diffuse dark matter halo as well as its stellar component.
The dark matter scale density $(\rho_0)$ of Antlia~2 is consistent with the result from \citet{2019MNRAS.488.2743T}.

We compare our results, especially for $\gamma$, with other studies.
Unfortunately, it is not possible to compare the inner slopes of dark matter density profiles in the UFDs.
However, there are several papers that have estimated their inner slopes.
\citet{2018MNRAS.481.5073E} adopted the mean density within 1.8 times half-light radius of a dSph and found that cuspy dark matter density profiles~($\gamma\sim1$) can reproduce the mean densities of  Canes Venatici~I, Canes Venatici~II, Coma Berenices,  Leo~T, Ursa~Major~II, Segue~1, Segue~2.
Their results are  roughly consistent with our results within the 68~\% confidence region.
On the other hand, \citet{2021A&A...651A..80Z} made an attempt to constrain the dark matter density profile of Eridanus~2 using CJAM~\citep{2008MNRAS.390...71C,2013MNRAS.436.2598W} and pyGravSphere~\citep{2017MNRAS.471.4541R,2020MNRAS.498..144G} with a generalized Hernquist profile.
From their analysis, the inner slope $\gamma$ is constrained as $\gamma>0.57$ (68~\% confidence region). This is also roughly consistent with our estimation within the 68~\% confidence region.

\subsubsection{Robustness of the estimated density profiles}
\label{sec:robustness}

To investigate the robustness of our results, especially regarding the inner slope of a dark matter density profile, $\gamma$, we perform the same MCMC fitting procedure~(see Section~\ref{sec:dataanalysis}) for the case of a wider range of prior for $\gamma$ than the fiducial parameter range ($0\leq\gamma\leq2$).
Namely, we adopt here the case of a flat prior over range $-2\leq\gamma^{\prime}\leq2$, and we impose $\gamma=0$ if $\gamma^{\prime}$ has a negative value and $\gamma=\gamma^{\prime}$ otherwise.
This is because the fiducial one might lead to a bias toward cuspy density profiles.
Using this new wide prior, we re-run the MCMC fitting procedure and estimate the dark matter density profiles for the representative eight galaxies.

Figure~\ref{fig:DMprofiles} shows the comparison of the inferred dark matter density profiles for the eight galaxies for the fiducial and wider prior ranges.
The colored solid line and shaded region in each panel show the results from the fiducial case, while the gray dashed lines denote the ones for the case of wider prior ranges.
Figure~\ref{fig:DMprofiles} shows that the estimated dark matter density profiles of most galaxies are largely affected by the new prior range. In particular, this prior makes their central densities clearly less dense.
Therefore, we note that most of the galaxies with small data volumes should be sensitive to the choice of priors.
This point is also highlighted in the mass-orbit modelings of dSphs~\citep[e.g.,][]{2015IAUS..311...16M}.

On the other hand, for galaxies with cuspy dark matter, such as Eridanus II, Segue 1 and Wilman 1, the inner slopes are not much changed with the wider range of priors, as expected.
The possible reason why these galaxies prefer cuspy dark matter density profiles comes from the following properties of their line-of-sight velocity dispersion profile in the central region.
Although the variations of $Q$ (dark halo shape) and $\beta_z$ (velocity anisotropy) give a similar effect on {\it entire} shape of line-of-sight velocity dispersion profiles~\citep[see][]{2008MNRAS.390...71C,2015ApJ...810...22H}, the inner slope of dark matter density profile, $\gamma$, can have an impact only on an {\it inner} velocity dispersion profile~\citep{2020ApJ...904...45H}. 
However, we bear in mind that the assumption of Gaussian line-of-sight velocity distribution causes a bias towards cuspy dark matter density profiles for dSphs with Plummer stellar density profile plus cored dark matter halo~\citep{2021MNRAS.501..978R}.

Note that cuspy dark matter density profiles in Eridanus~II, Segue~1 and Willman~1 come mainly from the kinematic sample in the inner part (especially within 10~pc) in our {\it unbinned} analysis. The unbinned analysis can trace the inner kinematic structures in galaxies, while the binned one might smear out such information, and thus may not provide such a constraint on inner slopes of dark matter density profile as the unbinned one~\citep[e.g.,][]{2014MNRAS.441.1584R}.

Finally, it is worth noting that even if a dSph is largely dominated by a gravitational potential by dark matter, the specific assumption of stellar density profiles can affect the inferred slope of a dark matter density profile at inner parts of a dSph~\citep[e.g.,][]{2009MNRAS.393L..50E,2012ApJ...755..145H}.
To address this issue, we take into account another functional form for the surface stellar density profile, namely, an exponential density profile\footnote{$\Sigma(x,y)\propto\exp(-m_\ast/b_\ast)$, where $m^2_\ast=x^2+y^2/q^{\prime2}$ and $b_\ast$ is scale radius. $(x,y)$ are the sky coordinates aligned with the major and minor axes, respectively.}.
We adopt this density profile to the same dynamical analysis as for a Plummer model. 
Taking the case of Segue~1 as an example, the constraints on the model parameters do not differ significantly, despite the differences in the selected stellar density profiles. This result may be attributed to Segue~1 being a poorly-sampled system, where statistical errors dominate the uncertainties in the model parameters.
For the case of Antlia~2 as another example, on the other hand, the inner slope of dark matter density estimated by an exponential stellar profile is slightly shallower ($\Delta\gamma=0.2$) than that estimated by a Plummer one, even though those values are roughly consistent within 68\% confidence level.
This is due to the fact that exponential three-dimensional profiles are somewhat steeper in the central regions compared to Plummer profiles. As a result, we confirm that there are systematic uncertainties, to some extent, for the inner slopes of dark matter density when different stellar density profiles are adopted.

\begin{figure}[t!!]
\begin{center}
 \includegraphics[width=\columnwidth]{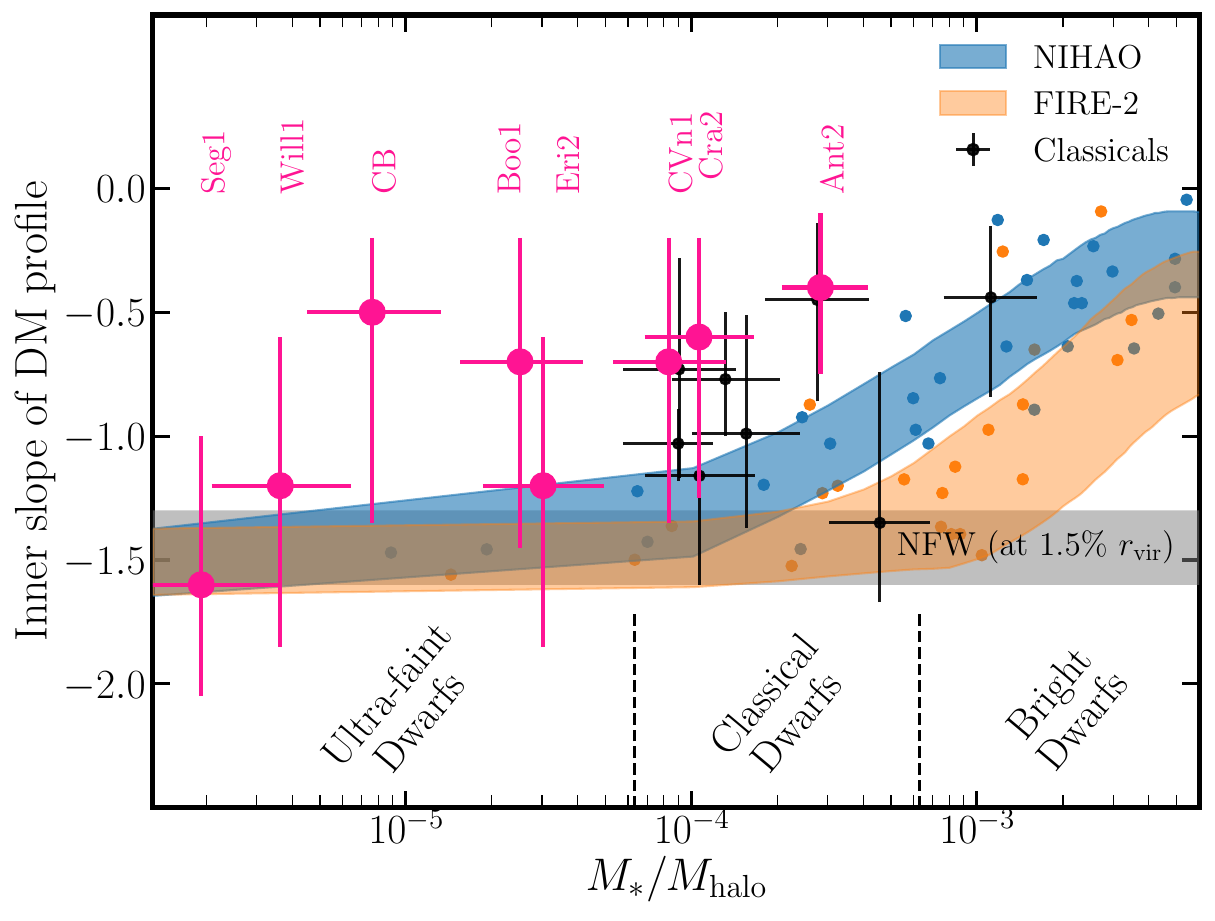}
\end{center}
\caption{The inner dark matter density slope at 1.5\% of the virial radius of the dark halo, $r_\mathrm{vir}$, as a function of the ratio between stellar and dark-halo masses, $M_\ast/M_\mathrm{halo}$, of the Galactic dSphs. The filled magenta circles with $1\sigma$~error bars are the results of the represented galaxies in this work, while the black filled ones are the results of the classical dSphs referred by \citet[][]{2020ApJ...904...45H}. 
The shaded gray band shows the expected range of dark matter profile slopes for NFW profiles as derived from dark-matter-only simulations~\citep{2016MNRAS.456.3542T}.
The blue and orange points are simulated satellites from NIHAO~\citep{2016MNRAS.456.3542T} and FIRE-2~\citep{2020MNRAS.497.2393L} hydrodynamical simulations, respectively.
The blue and orange shaded bands show the expected ranges from these simulated galaxies~(to guide the eye).}
\label{fig:gamma_MM}
\end{figure}

\subsubsection{Inner Dark Matter Density Slope versus Stellar-to-halo Mass Ratio}

Figure~\ref{fig:gamma_MM} shows the logarithmic slope of the dark matter density profile at 1.5\% of the virial radius of the dark halo as a function of the stellar mass-to-halo mass ratio, $M_\ast/M_\mathrm{halo}$. This is analogous to figure~6 of \citet[][originally following \citealt{2017ARA&A..55..343B}]{2020ApJ...904...45H}, but including several galaxies studied in our study.
The blue and orange dots and shaded bands depict the results from NIHAO~\citep{2016MNRAS.456.3542T} and FIRE-2~\citep{2020MNRAS.497.2393L} zoom-in hydrodynamical simulations, while the gray band shows the expected range of dark matter profile slopes for NFW as derived from dark matter-only simulations~\citep{2016MNRAS.456.3542T}.

To compute the stellar mass-halo mass ratios of the dwarf galaxies, we employ the self-consistent abundance matching model by~\citet{2013MNRAS.428.3121M} and adopt the stellar masses of most dSphs taken from the literature~(Table~\ref{table1}).
For several UFDs having no information about stellar masses, we calculate those of their UFDs by their luminosities by assuming a stellar mass-to-light ratio of $1.6M_\odot/L_\odot$, which is the median value for dSphs measured by~\citet{2008MNRAS.390.1453W}.
The black points with $1\sigma$ error bars are the results of classical dSphs estimated by \citet{2020ApJ...904...45H}, while the magenta ones are the largest eight galaxies among the sample in this work. 
According to the predictions from the simulations~(blue and orange bands), an inner slope of a dark matter density profile in the UFD regime~($M_\ast/M_\mathrm{halo}\lesssim10^{-4}$) should not be affected largely by baryonic feedback effects\redrev{, but we note from this figure that all of their inner dark matter slopes except one (Segue 1) are shallower than predicted by the hydrodynamical simulations.
}
On the other hand, there are large uncertainties in both the inner slopes and the stellar-to-halo mass ratios of the UFDs.
Thus, we cannot make a robust conclusion about whether the relation exists or not from the currently available data.

%%%%%%%%%%%%%%%%%%%%%%%%%%%%%%%%%%%%%%%%%%%%%%%%%%%%%
%%%%% Discussion %%%%%%%%%%%%%%%%%%%%%%%%%%%%%%%%%%%%%%%
%%%%%%%%%%%%%%%%%%%%%%%%%%%%%%%%%%%%%%%%%%%%%%%%%%%%%
\section{Discussion} \label{sec:Discussion}

\begin{figure}
\begin{center}
 \includegraphics[width=\columnwidth]{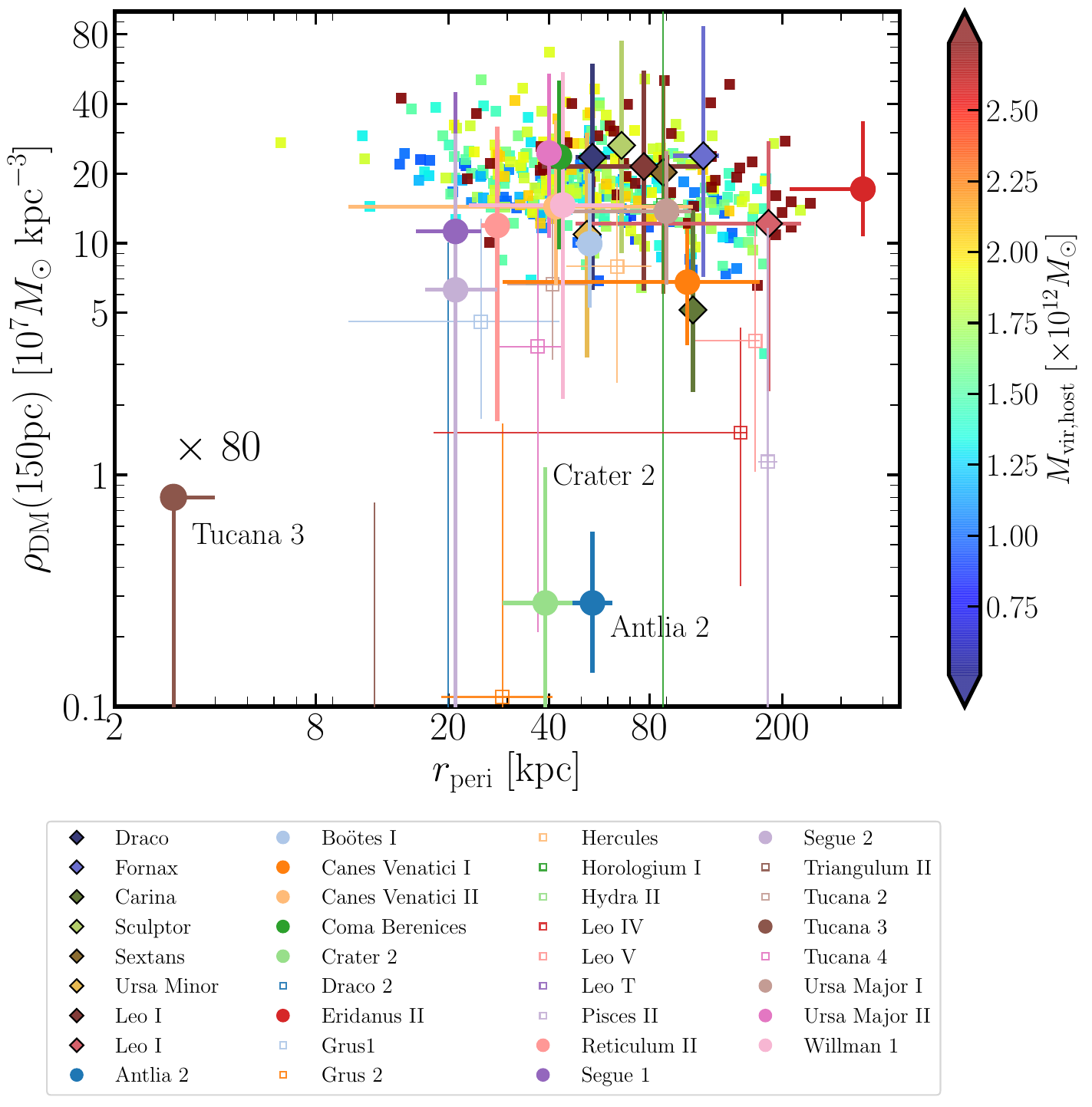}
\end{center}
\caption{Dark matter densities at $150$pc, $\rho_\mathrm{DM}(150\mathrm{pc})$, as a function of pericenter radii, $r_\mathrm{peri}$, of the dSphs. The small filled squares are the individual subhalos with $V_\mathrm{peak}>25$~km~s$^{-1}$ predicted from dark matter simulations~\citep{2021MNRAS.506.4210I}. The color scale indicates the virial masses of their host halos. The filled diamonds with error bars denote the results of the classical dSphs~\citep{2020ApJ...904...45H}.
The filled circles with error bars show the dwarf galaxies that have the number of kinematic sample grater that 20 stars, whilst open square ones are the galaxies have small sample less than 20 stars.
To plot Tucana~3 in this range, we multiply its density eightyfold.}
\label{fig:rho150}
\end{figure}

\subsection{Dark Matter densities at 150~pc}

\citet{2019MNRAS.484.1401R} introduced the local dark matter density at a common radius of 150~pc from the center of each dSph, $\rho_\mathrm{DM}(150\ \mathrm{pc})$, which is more robustly estimated than either the inner dark matter slope or its normalization $\rho_0$ in their spherical mass models.
They also argued that this density is enough different to divide the luminous dwarf galaxies into cusps or cores.
\citet{2019MNRAS.490..231K} found that $\rho_\mathrm{DM}(150\ \mathrm{pc})$ of the Galactic dSphs anti-correlates with their orbital pericenter distance, $r_\mathrm{peri}$, estimated by Gaia. 
In combination with the too-big-to-fail problem~\citep{2011MNRAS.415L..40B,2012MNRAS.422.1203B}, they proposed that this anti-correlation can provide a new incisive test of the nature of dark matter, especially for SIDM models~\citep[e.g.,][]{PhysRevD.101.063009,2021MNRAS.503..920C,2021arXiv210803243J}.
%For instance, several works indicated that SIDM models can reproduce this anti-correlation by both core-expansion and core-collapse phases, depending on the concentrations and pericenter distances of the subhalos~\citep[e.g.,][]{}.
%Moreover, using $\rho_\mathrm{DM}(150\ \mathrm{pc})$ of Draco,  \citet{2021arXiv210803243J} made an attempt to place constraints on the fundamental parameters of SIDM models.

Following these works, we calculate $\rho_\mathrm{DM}(150\ \mathrm{pc})$ along the major axis of the sample dwarf galaxies.
The calculated values are tabulated in the 11th column of Table~\ref{table2}.
Utilizing the part of these estimations, \citet{2022PhRvD.105b3016E} 
compared the $\rho_\mathrm{DM}(150\ \mathrm{pc})$-$r_\mathrm{peri}$ relation with that from the dark matter simulations based on $\Lambda$CDM and SIDM models with different (velocity independent) self-interaction cross sections~($\sigma/m=1,3$~cm$^2$~g$^{-1}$).
Although the uncertainties of $\rho_\mathrm{DM}(150\ \mathrm{pc})$ of UFDs are large, the SIDM scattering cross-section less than 3~cm$^2$~g$^{-1}$ could be favored.

We scrutinize the anti-correlation between $\rho_\mathrm{DM}(150\ \mathrm{pc})$-$r_\mathrm{peri}$ for all sample galaxies.
To do this with comparatively secure results, we separate the galaxies into two groups: the first group contains galaxies with at least 20 stars~(colored filled circles in Fig.~\ref{fig:rho150}), while the second group has fewer than 20 stars~(colored thin open circles in Fig.~\ref{fig:rho150})\footnote{This is motivated from the number of free parameters in our dynamical models. There are nine parameters to fit the observational quantities in this work.}. 
For the pericenter radius, we adopt the values estimated by \citet{2022A&A...657A..54B}, which analyzed the latest Gaia data~\citep[Gaia~EDR3,][]{2021A&A...649A...1G} assuming a Milky Way potential model with $8.8\times10^{11}M_\odot$\footnote{\redrev{We bear in mind that this pericenter estimation did not consider the dynamical effects of Large Magellanic Cloud~(LMC). It is possible that some of the dSphs have passed near LMC and thus have acquired angular momentum during their orbit leading to biases in their pericenters.}}.
Figure~\ref{fig:rho150} displays the relation between $\rho_\mathrm{DM}(150\ \mathrm{pc})$ and $r_\mathrm{peri}$.
The filled diamonds with error bars are the inferred $\rho_\mathrm{DM}(150\ \mathrm{pc})$ of the classical dSphs~\citep{2020ApJ...904...45H}, while the colored ones are the results of all sample dwarf galaxies in this work.
The dwarf galaxies are widely distributed in this plot, even though there are still large uncertainties.

Besides, Antlia~2, Crater~2, and Tucana~3 have very low dark matter densities, and thus they deviate significantly from the other ones. 
This deviation might be due to tidal effects. In particular, according to several studies of Tucana~3's orbital evolution based on {\it Gaia} satellites, this UFD is very likely to have traveled within 10~kpc from the Galactic center~\citep[e.g.,][]{2022A&A...657A..54B} and might be embedded in a tidal stream, which is stemmed from tidal disruption~\citep[e.g.,][]{2015ApJ...813..109D,2018MNRAS.481.3148E,2018ApJ...866...22L}. 
Therefore, this UFD might have been disturbed strongly by the gravitational potential of the Milky Way, and thus might have experienced a high mass loss.

We also compare dark matter subhalos predicted from dark matter-only simulations based on $\Lambda$CDM theory.
To create the subhalo sample, the simulated data we use and the selection procedure are the same as \citet{2020ApJ...904...45H}. We use the ROCKSTAR phase space halo/subhalo finder to select subhalos \citep{2013ApJ...762..109B}.
%To compute the $\rho_\mathrm{DM}(150\ \mathrm{pc})$ of the simulated dark subhalos, we use the scale density and radius of each subhalo, supposing a spherical NFW dark matter density profile.
In Figure~\ref{fig:rho150}, the small filled squares show the predicted subhalos associated with these Milky Way-sized dark matter host halos. 
These subhalos have $V_\mathrm{peak} >25$~km~s$^{-1}$, where $V_\mathrm{peak}$ is the maximum circular velocity of the subhalo over its assembly history.
The color scale of the squares indicates the virial masses of the host halos.
The simulated subhalos show a weak anti-correlation, which does not depend on their host halo masses.
On the other hand, although it does not seem that the Galactic satellites have such anti-correlation, these are reasonably consistent with the simulated subhalos.
As already discussed in the literature~\citep[e.g.,][]{2020ApJ...904...45H,2020arXiv201109482G}, the distribution could be explained by `survivor bias', which means denser central dark matter densities can survive even suffering from strong tidal effects at small pericenter and from repeated tides~\citep[e.g.,][]{2021MNRAS.505...18E}.

However, despite taking into account the tidal effects, the dark matter simulations have difficulty explaining the extremely low dark matter densities of Antlia~2, Crater~2, and Tucana~3.
This discrepancy was already argued by several works \citep[e.g.,][]{2020arXiv200606681S,2021ApJ...921...32J,2022MNRAS.512.5247B}.
Obviously, we should bear in mind that our dynamical analysis is based on dynamical equilibrium, while such diffuse systems can deviate from the dynamical equilibrium.
Thus, our resultant dark matter densities could have systematic uncertainties stemmed from this assumption. 
In addition, it can be difficult for commonly-used subhalo finders~\citep[e.g.,][]{2013ApJ...762..109B} to detect the substructures that have very low dark matter densities, because of low-contrast dark matter densities between a subhalo and the background~(i.e., a host halo).
Therefore, it should be noted that such diffuse subhalos can be easily overlooked by the subhalo finders.

\begin{figure}
\begin{center}
 \includegraphics[width=\columnwidth]{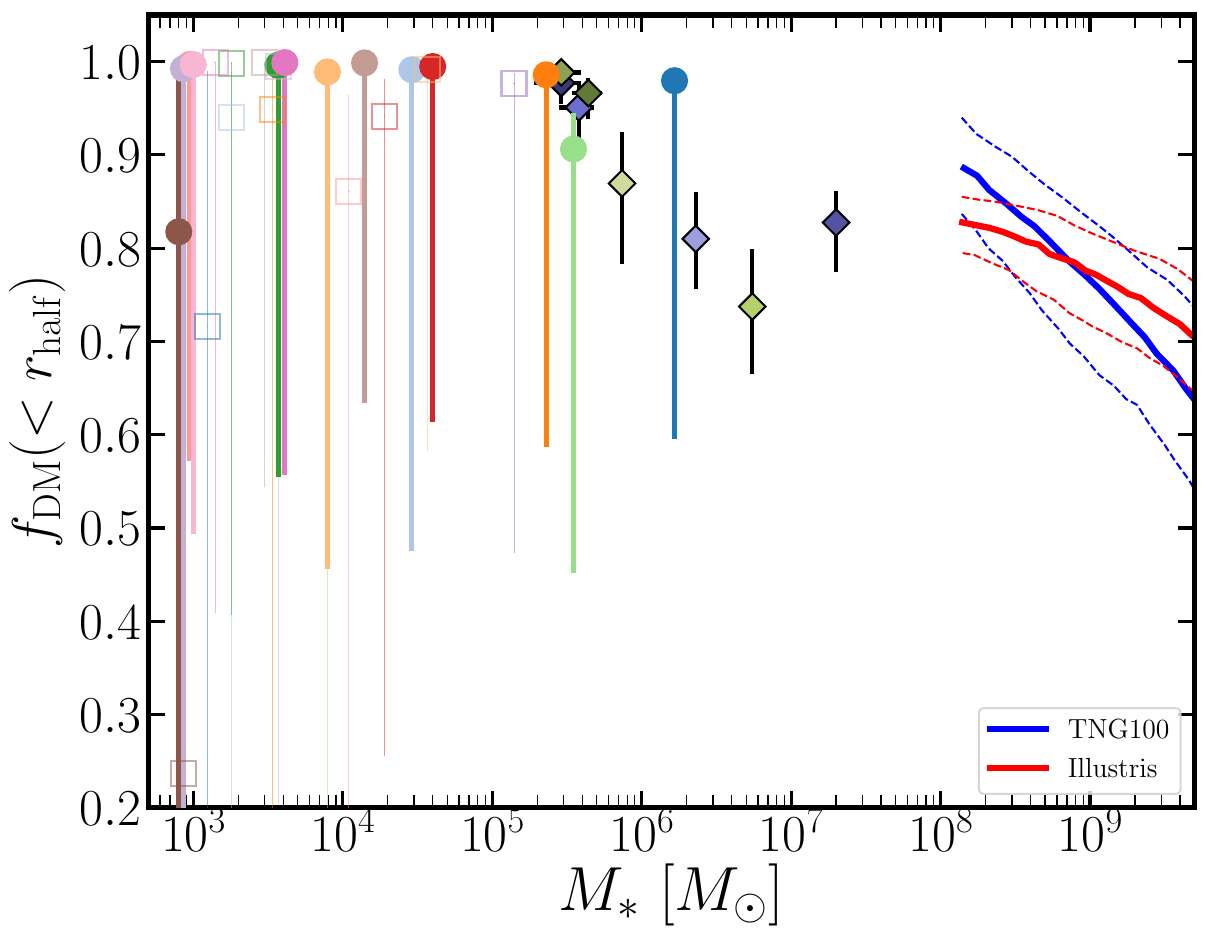}
\end{center}
\caption{\redrev{Mass fraction of dark matter in the dSphs within the half-light radius~($r_\mathrm{half}$) as a function of galaxy stellar mass. The colors and symbols for the dSphs are the same as Fig.~\ref{fig:rho150}.
The TNG100 and Illustris simulated data \citep[taken from][]{2018MNRAS.481.1950L} are shown in blue and red, respectively. The solid and dashed lines denotes the median and 68~per~cent of their data.}}
\label{fig:DMfraction}
\end{figure}

%#########################################
%#########################################
\subsection{\redrev{Dark matter fractions}}
%#########################################
%#########################################
\redrev{\citet{2018MNRAS.481.1950L} investigated the relative content of dark matter and baryonic mass using the IllustrisTNG \citep[TNG,][]{2018MNRAS.475..676S} and Illustris \citep{2014MNRAS.444.1518V} simulations, in order to make theoretical predictions about the relationship between a galaxy's total mass and its stellar component.
They selected simulated galaxies with $M_\ast\sim10^8-10^{12}M_\odot$ at $z=0$, and then computed their dark matter fraction (DMF), $f_\mathrm{DM}=M_\mathrm{DM}/(M_\mathrm{DM}+M_\ast)$ within several kinds of radii as a function of galaxy stellar mass~(e.g., half-light radius $r_\mathrm{half}$).
They predicted that $f_\mathrm{DM}( < r_\mathrm{half})$ is around 0.8 at $M_\ast\sim10^8M_\odot$ and gradually decreases toward 0.5-0.6 with increasing stellar mass at $M_\ast\sim10^{10}M_\odot$ for the case of DMF within $r_\mathrm{half}$,.}

\redrev{Following this work, we calculate $f_\mathrm{DM}( < r_\mathrm{half})$ of the sample dSphs. Figure~\ref{fig:DMfraction} shows the $f_\mathrm{DM}( < r_\mathrm{half})$ as a function of $M_\ast$. The symbols are the same as those in Figure~\ref{fig:rho150}. The TNG100 and Illustris simulated data \citep[taken from][]{2018MNRAS.481.1950L} are shown in blue and red, respectively. The solid and dashed lines denote the median and 68\% of their data.
While most of low-mass dSphs have $f_\mathrm{DM}\sim 1.0$, the dark matter fractions of massive ones decrease to $f_\mathrm{DM}\sim 0.7-0.8$, but there are large uncertainties.
On the other hand, due to resolution limits, the simulation results cannot reach low-mass galaxies less than $M_\ast\simeq10^8M_\odot$.
In order to inspect whether the simulated fractions are consistent with the observed ones, it is necessary to perform much more high-resolution cosmological simulations.}

%#########################################
%#########################################
\subsection{Dynamical mass-to-light ratios}
%#########################################
%#########################################

As a by-product of the fitting results, we estimate the dynamical mass-to-light ratios~(MLs) within the half-light radii of the dSphs, which we write $(M_\mathrm{dyn}/L)_{r_\mathrm{half}}$.
This is motivated by the increasing mass-to-light ratio for decreasing dwarf luminosity~\citep{1998ARA&A..36..435M,2010MNRAS.406.1220W,2012AJ....144....4M}, whose mass-to-light ratios increase remarkably with decreasing luminosity.
We therefore calculate $(M_\mathrm{dyn}/L)_{r_\mathrm{half}}$ of our sample dwarf galaxies as well as the classical dSphs to examine whether the scaling law still sustain even with considering the non-sphericity of their dark matter halos and the latest data sample.

\begin{figure*}[t!!]
\begin{center}
 \includegraphics[width=\textwidth]{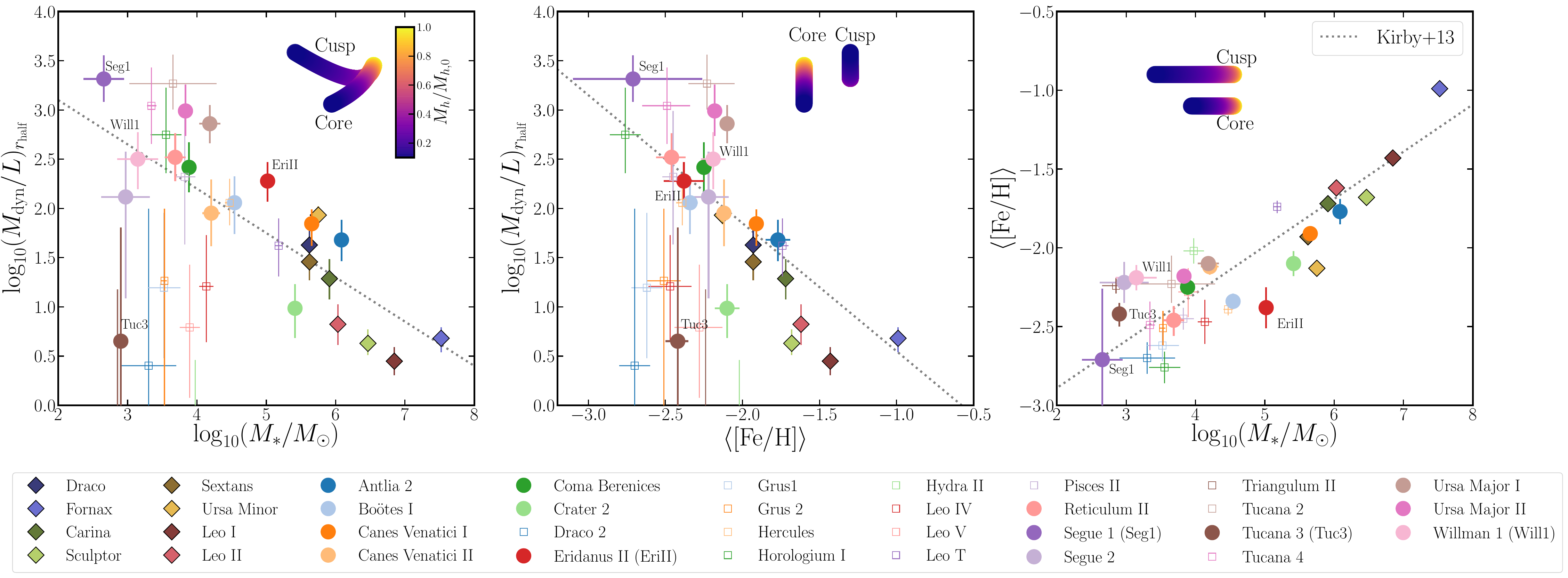}
\end{center}
\caption{{\it Left}: The dynamical mass-to-light ratio within a half-light radius, $(M_\mathrm{dyn}/L)_{r_\mathrm{half}}$, as a function of stellar mass for the dwarf galaxies. The gray dotted line shows $\log_{10}(M_\mathrm{dyn}/L)_{r_\mathrm{half}}\propto -0.4\log_{10}M_\ast$ suggested by NFW dark matter halos.
{\it Middle}: $(M_\mathrm{dyn}/L)_{r_\mathrm{half}}$ as a function of averaged metallicity. Whereas the dotted line denotes $\log_{10}(M_\mathrm{dyn}/L)_{r_\mathrm{half}} \propto -1.3\langle\mathrm{[Fe/H]}\rangle$ induced from an NFW dark matter halo and MZR (the dotted lines in the right and middle panels).
{\it Right}: The stellar mass versus averaged metallicity. The gray dashed line in this panel is the universal MZR~($\langle\mathrm{[Fe/H]}\rangle\propto0.3\log_{10}M_\ast$), taken from~\citet{2013ApJ...770...16K}.
The colored thick lines in each panel correspond to the tidal evolutionary track based on~\citet[][see also~\citealt{2008ApJ...673..226P,2018MNRAS.476.3816F}]{2015MNRAS.449L..46E}.
The color scale indicates the subhalo total mass fraction that is retained against tidal stripping. An initial one~(yellow) means the fraction before the halo were affected by tidal effects.
In the middle panel, the yellow color for cuspy case seems to be disappeared. The yellow is hidden by the blue, because the evolution causes a down then up movement in $(M_\mathrm{dyn}/L)_{r_\mathrm{half}}$, as can be seen in the left panel.}
Here, we suppose that the metallicities in the dwarf galaxies are unaffected by tidal effects~(i.e., they shift only vertically (horizontally) in middle (right) panel).
\label{fig:MLs3fig}
\end{figure*}

The dynamical MLs we estimated are shown in the 10th column of Table~\ref{table2}, and Figure~\ref{fig:MLs3fig} shows the dynamical mass-to-light ratios of all dSphs in our sample, as a function of stellar mass~(left panel) and of averaged metallicity~(middle panel).
Here, we focus only on the filled symbols, which correspond to the galaxies having large kinematic sample sizes larger than 20 stars.
In the left panel of Figure~\ref{fig:MLs3fig}, the dotted line indicates $\log_{10}(M_\mathrm{dyn}/L)_{r_\mathrm{half}}\propto -0.4\log_{10}M_\ast$, suggested by \redrev{the scaling relations found by \citet{2018MNRAS.476.3816F} in their hydrodynamical simulations: $M_\ast\propto r^{7/2}$, and  $V_{\rm{circ}}(r_{\rm{half}})\propto r^{1/2}$ as well as $M_{\rm dyn} \propto r_{\rm{half}}\sigma_{\rm los}^2$ and $M_\ast\propto L_\ast$}.
%an NFW dark matter density profile, which this relation is consistent with the  satellites computed from $\Lambda$CDM based cosmological hydrodynamical simulations of the Local Group~\citep{2018MNRAS.476.3816F}.

Figure~\ref{fig:MLs3fig} shows that most of the Galactic dSphs agree with the suggested relation.
However, there is non-negligible scatter in the MLs, not seen 
in the simulations~\citep[see Figure~7 in][]{2018MNRAS.476.3816F}, especially for the simulated galaxies with low-MLs. 
There are several explanations for this discrepancy:
i) the simulations might not resolve dark matter densities at UFD-mass scales and with very diffuse galaxies, ii) they might not be able to identify the galaxies that were strongly stirred by the tidal effects, and iii) the simulated less-massive galaxies could maintain cuspy dark matter densities even considering stellar feedback effects or iv) biases in the kinematic mass modeling and/or underestimated uncertainties on $M_{\rm{dyn}}$.
To verify these possibilities, high-resolution hydrodynamical simulations of Milky Way analogs are required. However, this is beyond the scope of the present article.

Meanwhile, we confirm that the dynamical ML might anti-correlate with the averaged metallicity even at the low-metallicity end, as shown in the right panel of Figure~\ref{fig:MLs3fig}.
The dotted line in this panel shows the relation $\log_{10}(M_\mathrm{dyn}/L)_{r_\mathrm{half}}\propto -1.3\langle\mathrm{[Fe/H]}\rangle$ derived by NFW dark matter halo ($\log_{10}(M_\mathrm{dyn}/L)_{r_\mathrm{half}}\propto -0.4\log_{10}M_\ast$) and the universal MZR~\citep[][]{2013ApJ...770...16K}, which indicates $\langle\mathrm{[Fe/H]}\rangle\propto0.3\log_{10}M_\ast$.
From this panel, the observed dSphs are in line with the relation, even though a large scatter still exists.

%To characterize this anti-correlation quantitatively, we employ a least-squares fitting method to determine $(M_\mathrm{dyn}/L)_{r_\mathrm{half}}$ as a function of averaged metallicity, $\langle\mathrm{[Fe/H]}\rangle$, and we find $\log_{10}(M_\mathrm{dyn}/L)_{r_\mathrm{half}} \propto -2.3\langle\mathrm{[Fe/H]}\rangle$.

The scatter in ML with stellar mass and with metallicity, shown in Figure~\ref{fig:MLs3fig}, might stem from tidal effects, even though non-sphericity of the dark matter halo and deviation from an NFW density profile could also be possible reasons for the scatters.
In the following section, we discuss the tidal effects on the dynamical ML relations, taking into account the emperical tidal evolution models.

\subsection{Tidal evolution of the dwarf satellites}
To examine the tidal effects on $(M_\mathrm{dyn}/L)_{r_\mathrm{half}}$ and $M_\ast$, we estimate the evolution of these based on a simple empirical formula calculated by \citet{2015MNRAS.449L..46E}.
This formula, as known as {\it tidal evolutionary tracks}, describes parametrically the tidal evolution of the half-light radius~$r_h$, the luminosity averaged velocity dispersion $\sigma_h$ within $r_h$, the total stellar mass $M_\ast$, and the total mass~$M_h$ enclosed within $r_h$, which are all normalized to their initial values under the assumption that the stellar density profile follows that of the Plummer sphere.
\citet{2015MNRAS.449L..46E} performed high-resolution $N$-body simulations for the cases of cored and cuspy dark matter subhalos and searched the empirical fitting parameters to their tidal tracks proposed by \citet{2008ApJ...673..226P}.
The fitting formula is written as
\begin{eqnarray}
f(x) = \frac{2^\alpha x^\beta}{(1+x)^\alpha} \,\,\,\,\,\, \mathrm{where} \,\,\,\, x=M_h/M_{h,0}.
\end{eqnarray}
Here, $x$ is the total mass $M_h$ that remains bound within the initial stellar half-mass radius of the satellite, in units of the pre-stripping value, $M_{h,0}$.
$\alpha$ and $\beta$ are the fitting parameters.
The best-fit values for both core and cusp cases are tabulated in Table~1 in \citet{2015MNRAS.449L..46E}\footnote{In this paper, we adopt the best-fit values as follows. {\it Core cases}: $(\alpha,\beta)=(1.63,0.03)$ for $r_h/r_{h,0}$ and $(\alpha,\beta)=(0.82,0.82)$ for $M_\ast/M_{\ast,0}$, 
{\it Cusp cases}: $(\alpha,\beta)=(1.22,0.33)$ for $r_h/r_{h,0}$ and $(\alpha,\beta)=(3.57,2.06)$ for $M_\ast/M_{\ast,0}$}.
In addition, the empirical formula with best-fit values is in accord with the results from APOSTLE simulations, which are cosmological hydrodynamical simulations of the Local Group~\citep{2018MNRAS.476.3816F}.

Using the tidal evolutionary tracks of $M_\ast$ and $M_h$, relative to their unstripped values, we calculate the tidal evolution of $(M_\mathrm{dyn}/L)_{r_\mathrm{half}}$.
The estimated tidal evolution of $(M_\mathrm{dyn}/L)_{r_\mathrm{half}}$ are plotted on the $(M_\mathrm{dyn}/L)_{r_\mathrm{half}}-M_\ast$ and $(M_\mathrm{dyn}/L)_{r_\mathrm{half}}-\langle\mathrm{[Fe/H]}\rangle$ planes of Figure~\ref{fig:MLs3fig}.
The thick colored lines in each panel display the evolutionary tracks for the cases of cuspy and cored dark matter subhalos.
The color bar indicates \redrev{the subhalo total mass fraction that is retained against tidal stripping}.
The yellow color indicates $M_h=M_{h,0}$, which means the dark halo is pre-infalling.  
Here, we assume that metallicities in the dwarf galaxies are unaffected by tidal effects~(i.e., they shift only horizontally (vertically) in the middle (right) panel in Figure~\ref{fig:MLs3fig})\footnote{If a negative metallicity gradient in a dwarf galaxy exists, the metallicity could be enriched by tides because metal-poor stars in outer parts of a galaxy are preferentially stripped.}.

The left and middle panels of Figure~\ref{fig:MLs3fig} indicate that for subhalos with cuspy dark matter, tidal stripping does not alter much either scaling relation for $(M_\mathrm{dyn}/L)_{r_\mathrm{half}}-M_\ast$ and $(M_\mathrm{dyn}/L)_{r_\mathrm{half}}-\langle\mathrm{[Fe/H]}\rangle$.
Instead, for subhalos with constant density cores, $(M_\mathrm{dyn}/L)_{r_\mathrm{half}}$ is changed more clearly.
In the right panel, on the other hand, we find that a stellar mass embedded in a cuspy dark matter halo is decreased prominently by tidal stripping.
These apparent counter-intuitive results are already discussed by \citet[][see also~\citealt{2008ApJ...673..226P}]{2015MNRAS.449L..46E}.
They interpreted that most of particles tagged as `stars' in a cored dark matter halo are associated to the most-bound dark matter particles, whilst a small fraction of stars are associated to dark matter particles with high binding energy in a cusped dark matter halo.
In short, Figure~\ref{fig:MLs3fig} indicates that the tidal evolution of dwarf galaxies with cuspy halos does not affect the scaling relations of ML with stellar mass and with metallicity, but does affect the MZR.
%from the three panels in Figure~\ref{fig:MLs3fig}, dwarf galaxies associated with cuspy dark matter halos are tidally evolved so that they do not change two scaling relations~(i.e., $(M_\mathrm{dyn}/L)_{r_\mathrm{half}}-M_\ast$ and $(M_\mathrm{dyn}/L)_{r_\mathrm{half}}-\langle\mathrm{[Fe/H]}\rangle$) but do change MZR.
These results are consistent with the results from cosmological hydrodynamical simulations~\citep{2018MNRAS.476.3816F}.
Interestingly, the presence of negative metallicity gradients can potentially amplify the above results.
Moreover, the results imply that the large scatter of metallicity at the faint-end of the MZR can be explained by tidal stripping of less-massive satellites associated with cuspy dark matter halos\redrev{, even though this scatter can also be caused by the lack of data to properly measure stellar mass and metallicity}.

Indeed, the results of Segue~1 and Willman~1, which favor cuspy dark matter density profiles, can be interpreted by the tidal evolution within their finite uncertainties.
The stellar and dark matter halo masses of their progenitors would have been more massive. After the accretion onto the Milky Way, the tidal effects can decrease their stellar masses, but the MLs are not much altered.
However, although Eridanus~II also appears to have a cuspy dark matter halo, it might not be evolved along the tracks. In particular, the deviation from the MZR cannot be explained by tidal effects.
Besides, the pericenter distance of Eridanus~II may be very large~($r_\mathrm{peri}\sim350$~kpc estimated by using Gaia EDR3, e.g., \citealt{2022A&A...657A..54B}), and thus this galaxy should not suffer from tidal forces.
Therefore, the chemo-dynamical evolution of this galaxy could be interpreted by the scatter induced from the merging and star formation histories, baryon feedback, and metallicity gradient of isolated systems~\citep[e.g.,][]{2016ApJ...827L..23W,2021MNRAS.501.5121M}.

On the other hand, several galaxies, especially Tucana~3, deviate significantly from the scaling relations in Figure~\ref{fig:MLs3fig} except for the MZR.
As mentioned above, Tucana~3 might have been disturbed strongly by the tidal force because at pericenter, it is located very close to the center of Milky Way (3-4~kpc according to \citealt{2022A&A...657A..54B}).
Thus, the departure of Tucana~3 from the $(M_\mathrm{dyn}/L)_{r_\mathrm{half}} - M_\ast$ on one hand and $(M_\mathrm{dyn}/L)_{r_\mathrm{half}} - \langle\mathrm{[Fe/H]}\rangle$ on the other hand, combined with its normal mass metallicity relation indicates that Tucana~3 has a cored dark matter halo.
Even though several UFDs~(such as Draco~II, Segue~1, Ursa~Major~I, Willman~1, etc.) can also be considered as tidally stirred~\citep[e.g.,][]{2022A&A...657A..54B}, they might not have suffered stronger tidal effects than Tucana~3.
However, we bear in mind that there is a considerable uncertainty on the $(M_\mathrm{dyn}/L)_{r_\mathrm{half}}$ estimation of Tucana~3 due to the small number (26) of available velocities and the dynamical non-equilibrium provoked by strong tidal effects.

%Moreover, it is not strong enough observational evidence to validate tidal features.

%%%%%%%%%%%%%%%%%%%%%%%%%%%%%%%%%%%%%%%%%%%%%%%%%%%%%
%%%%% Conclusion %%%%%%%%%%%%%%%%%%%%%%%%%%%%%%%%%%%%%%%
%%%%%%%%%%%%%%%%%%%%%%%%%%%%%%%%%%%%%%%%%%%%%%%%%%%%%
\section{Summary and Conclusion} \label{sec:Conclusion}
In this paper, we investigate dark matter halo properties of the Galactic ultra-faint dwarf and diffuse galaxies utilizing non-spherical dynamical mass models based on axisymmetric Jeans equations. 

Applying our non-spherical mass models to the latest kinematic data of the 25~UFDs and 2 diffuse galaxies in the Milky Way, we estimate their dark matter density profiles by marginalizing posteriors of non-spherical dark matter halo parameters.
We find (Figure~\ref{fig:DMprofiles}) that most of these galaxies have large uncertainties on the inferred dark matter density profiles especially for their inner density slopes, which are largely affected by the choice of the range of priors.
On the other hand, the dark matter density profiles of Eridanus~II, Segue~I, and Willman~1 are are not much affected by the choice of priors, and their inner slopes appear to be cuspy.

Utilizing their inferred central dark matter density profiles, we compare them with the predictions from recent cosmological zoom-in hydrodynamical simulations on the relation between the inner slope of a dark matter density profile and stellar mass-to-halo mass ratio in Figure~\ref{fig:gamma_MM}.
Although some of our sample galaxies are roughly consistent with the predictions from recent numerical simulations, there are large uncertainties in both their inner slopes and stellar-to-halo mass ratios.
Thus, we cannot make a conclusion about whether the relation exists or not from the current available data.

We scrutinize the anti-correlation between a dark matter density at a radius of 150~pc and the orbital pericenter distance of a dwarf satellite highlighted by~\citet{2019MNRAS.490..231K} and compare it with simulated dark matter subhalos based on $\Lambda$CDM models~(Figure~\ref{fig:rho150}).
Although it does not seem that the dwarf galaxies studied here have this anti-correlation, these are roughly consistent with the simulated dark matter subhalos.
Nevertheless, Antlia~2, Crater~2, and Tucana~3 have very low central dark matter densities, thus deviating significantly from the anti-correlation.
Even after taking into account tidal effects, dark matter simulations have difficulty in explaining such large deviations from the density-mass relation.
However, we should bear in mind that commonly-used subhalo finders are not very able to detect substructures that have very low dark matter densities.

Following several simulation works such as the IllustrisTNG and Illustris, we calculate dark matter fraction within a half-light radius, $f_\mathrm{DM}(< r_\mathrm{half})$ of the sample dSphs.
While most of low-mass dSphs have $f_\mathrm{DM}\sim 1.0$, the dark matter fractions of massive ones decrease to $f_\mathrm{DM}\sim 0.7-0.8$, but there are large uncertainties.
On the other hand, due to resolution limits, the simulation results cannot reach low-mass galaxies less than $M_\ast\simeq10^8M_\odot$.
In order to inspect whether the simulated fractions are consistent with the observed ones, it is necessary to perform much higher resolution cosmological simulations.

We also estimate the dynamical mass-to-light ratios~(MLs) within the half-light radii of our sample dwarf galaxies~$(M_\mathrm{dyn}/L)_{r_\mathrm{half}}$.
%and investigate the relations of MLs-stellar mass and MLs-averaged metallicities.
We confirm that most Galactic dwarf satellites agree with the scaling relation between the dynamical MLs and stellar mass proposed by an NFW dark matter density profile as well as the relation between the MLs and the averaged metallicities at low-metallicity end, even though there are large uncertainties in MLs~(Figure~\ref{fig:MLs3fig}).

Using these scaling relations for MLs and the universal MZR, we discuss the tidal evolution of the dwarf satellites based on empirical {\it tidal evolutionary tracks}.
From the evolutionary tracks, if these satellites have cuspy dark matter halos, tidal stripping does not alter much the dynamical mass-to-light ratios.
On the other hand, the MLs of the satellites having constant density cores might be changed more clearly.
In light of the MZR, tidal stripping might be capable of decreasing a stellar mass.
We suggest that these results imply that the
large scatter of metallicity at the faint-end of the MZR can be explained by tidal stripping of less-massive satellites associated with cuspy dark matter halos.

To ensure our conclusions, especially for tidal evolution for the satellites, it is necessary to perform high-resolution hydrodynamical simulations in Milky Way analogs so that less-massive dwarf satellites can also be resolved~\citep[e.g.,][]{2021ApJ...906...96A,2021MNRAS.507.4953G, 2022MNRAS.517.4856H}.
In this regard, using high-resolution simulations in our group, we will make an attempt to investigate more details of the chemo-dynamical evolution of the dwarf satellites in a forthcoming paper.
It is also important to estimate more accurately the dark matter density profiles of the Galactic dwarf satellites.
To this end, incorporating higher-order moments of their velocity distributions can help to determine uniquely the kinematic properties of their stellar systems and to mitigate a degeneracy between dark matter density and stellar anisotropy.
Furthermore, the next generation wide-field spectroscopic surveys with the Subaru Prime Focus Spectrograph~\citep{2014PASJ...66R...1T,2016SPIE.9908E..1MT} and high-precision spectroscopy attached on the Thirty Meter Telescope~\citep{2016SPIE.9908E..1VS} will enable us to obtain unprecedented data for stellar kinematics and chemical abundances for the satellites to tighter constrain their dark matter distributions.

\section*{Acknowledgements}
We are grateful to the referee for her/his careful reading of our paper and thoughtful comments.
%We would like to give special thanks to xxxxxx for useful discussions.
This work was supported in part by the MEXT Grant-in-Aid for Scientific Research  (No.~JP20H01895, JP21K13909, JP21H05447 and JP23H04009 for K.H., No.~JP22KJ0157, JP20K14532, JP21H04499, JP21K03614 and JP22H01259 for~Y.H., No.~JP18H05437, and JP21H05448 for~M.C., No.~JP20H05245 and JP21H01122 for~T.I.).
Numerical computations were partially carried out on Aterui II supercomputer at
Center for Computational Astrophysics, CfCA, of National Astronomical
Observatory of Japan. Y.H. and T.I. have been supported by JICFuS 
and MEXT as ``Program for Promoting Research on the Supercomputer
Fugaku'' (Structure and Evolution of the Universe Unraveled by Fusion of Simulation and AI, Grant No. JPMXP1020230406). 
T. I. has been supported by IAAR Research Support Program, Chiba University, Japan.
This work was supported in part by the National Science Foundation under Grant No. PHY-1430152 (JINA Center for the Evolution of the Elements).

Software: NumPy~\citep{2020Natur.585..357H}, Matplotlib~\citep{4160265}, SciPy~\citep{2020NatMe..17..261V}, corner~\citep{corner}.
%%%%%%%%%%%%%%%%%%%%%%%%%%%%%%%%%%%%%%%%%%%%%%%%%%%%%
%%%%% Appendix %%%%%%%%%%%%%%%%%%%%%%%%%%%%%%%%%%%%%%%
%%%%%%%%%%%%%%%%%%%%%%%%%%%%%%%%%%%%%%%%%%%%%%%%%%%%%
\appendix
\restartappendixnumbering

\section{Tabulated model parameter constraints} \label{sec:appA}
In this appendix, we present the constraints on the individual model parameters for all sample dwarf galaxies.
Table~\ref{table2} shows the estimated parameters~$(Q,b_\mathrm{halo},\rho_0,\beta_z,\alpha,\beta,\gamma,i)$ with the error values indicate the 68\% confidence intervals computed from the posterior distribution functions of the parameters.
We also show the dynamical mass-to-light ratios within their half-light radii, the dark matter densities at 150~pc, and the Bayes factors for the addition of a free halo axis ratio.

%%% Table 2 %%%
 \setlength{\tabcolsep}{1.5pt}
\begin{table*}
\centering
	\caption{Parameter constraints for the UFDs. Errors correspond to the $1\sigma$ range of our analysis. \redrev{The bold names of galaxies are those which have the number of kinematic sample greater than 20 stars.}}
	\label{table2}
	%\scalebox{0.77}[0.85]{
\begin{center}
	\begin{tabular}{lcccccccccccc}
 	% four columns, alignment for each
		\hline\hline
Object  & $Q$ & $\log(b_{\rm halo})$ & $\log$($\rho_0)$   & $-\log(1-\beta_z)$ & $\alpha$  & $\beta$  & $\gamma$ & $i$ & $(M_\mathrm{dyn}/L)_{r_\mathrm{half}}$  & $\log\rho_\mathrm{DM}(150\mathrm{pc})$ & 2$\ln({\rm BF_1})^{\rm a}$ & 2$\ln({\rm BF_2})^{\rm a}$ \\
        &     &            [pc]           & [$M_{\odot}$~pc$^{-3}$] &                         &           &          &          & [deg] & [$M_\odot / L_\odot$] & [$M_\odot$ kpc$^{-3}$]  &\\
		\hline
{\bf Antlia 2}
& $1.1^{+0.6}_{-0.6}$ & $4.5^{+0.4}_{-0.3}$ & $-3.4^{+0.5}_{-0.4}$ & $0.3^{+0.3}_{-0.3}$ & $2.0^{+0.7}_{-0.7}$ & $6.2^{+2.0}_{-2.2}$ & $0.4^{+0.2}_{-0.3}$ & $69.7^{+11.2}_{-13.3}$ & 
$47.9_{-18.7}^{+28.8}$ & $6.4_{-0.3}^{+0.3}$ & 1.0 & 0.5 \\ 

{\bf Bo$\ddot{\mathrm{o}}$tes I}
& $1.1^{+0.6}_{-0.6}$ & $3.7^{+0.9}_{-0.8}$ & $-1.8^{+1.2}_{-0.9}$ & $0.1^{+0.5}_{-0.4}$ & $1.8^{+0.8}_{-0.8}$ & $6.3^{+2.1}_{-2.2}$ & $0.7^{+0.5}_{-0.6}$ & $73.2^{+11.7}_{-10.9}$ & 
$114_{-59}^{+98}$& $8.0_{-0.3}^{+0.3}$ & 0.0  & -0.4 \\ 

{\bf Canes Venatici I}
& $1.2^{+0.6}_{-0.5}$ & $4.0^{+0.8}_{-0.7}$ & $-2.2^{+1.1}_{-0.7}$ & $0.4^{+0.2}_{-0.3}$ & $1.9^{+0.8}_{-0.7}$ & $6.3^{+2.1}_{-2.2}$ & $0.7^{+0.4}_{-0.5}$ & $74.2^{+11.2}_{-10.1}$ & 
$69.9_{-28.3}^{+27.9}$ & $7.9_{-0.3}^{+0.2}$ & -0.1 & 0.0 \\ 

{\bf Canes Venatici II}
& $1.1^{+0.6}_{-0.6}$ & $3.7^{+0.9}_{-0.8}$ & $-1.8^{+1.4}_{-1.0}$ & $-0.2^{+0.5}_{-0.6}$ & $1.8^{+0.8}_{-0.8}$ & $6.3^{+2.1}_{-2.2}$ & $0.8^{+0.5}_{-0.5}$ & $73.8^{+8.9}_{-10.3}$ &
$89.5_{-48.3}^{+107.4}$ & $8.2_{-0.3}^{+0.4}$ & 0.0 & -0.1 \\ 

{\bf Coma Berenices}
& $1.2^{+0.6}_{-0.5}$ & $3.6^{+1.0}_{-0.9}$ & $-1.1^{+1.4}_{-0.7}$ & $-0.2^{+0.5}_{-0.5}$ & $1.9^{+0.8}_{-0.7}$ & $6.3^{+2.1}_{-2.2}$ & $0.5^{+0.4}_{-0.4}$ & $71.0^{+12.1}_{-12.3}$ & 
$261_{-116}^{+209}$& $8.4_{-0.4}^{+0.3}$ & 0.0 & 0.0 \\ 

{\bf Crater~2}
& $0.7^{+0.4}_{-0.7}$ & $3.6^{+0.6}_{-0.6}$ & $-3.6^{+0.8}_{-1.0}$ & $0.8^{+0.2}_{-0.1}$ & $1.6^{+0.8}_{-0.9}$ & $6.1^{+2.1}_{-2.4}$ & $0.9^{+0.5}_{-0.5}$ & $88.0^{+1.5}_{-1.3}$ & 
$9.1_{-4.0}^{+5.8}$ & $6.8_{-0.4}^{+0.4}$ & 0.3 & 0.0 \\ 

Draco 2
& $1.1^{+0.6}_{-0.6}$ & $2.1^{+1.4}_{-1.8}$ & $-1.9^{+2.0}_{-2.5}$ & $-0.2^{+0.5}_{-0.5}$ & $1.7^{+0.8}_{-0.8}$ & $6.4^{+2.1}_{-2.2}$ & $0.9^{+0.6}_{-0.7}$ & $62.4^{+15.8}_{-17.8}$ &
$2.5_{-2.5}^{+96.9}$& $5.1_{-6.4}^{+2.9}$ & -0.1 & -0.1 \\ 

{\bf Eridanus II}
& $1.2^{+0.6}_{-0.5}$ & $3.8^{+0.8}_{-0.7}$ & $-2.6^{+1.2}_{-1.3}$ & $0.5^{+0.2}_{-0.2}$ & $1.9^{+0.9}_{-0.8}$ & $6.3^{+2.1}_{-2.3}$ & $1.2^{+0.6}_{-0.4}$ & $71.2^{+11.6}_{-12.4}$ & 
$190_{-73}^{+107}$& $8.2_{-0.2}^{+0.3}$ & -0.1 & 0.4 \\ 

Grus 1
& $1.1^{+0.6}_{-0.6}$ & $3.6^{+0.9}_{-0.9}$ & $-2.0^{+1.2}_{-1.0}$ & $0.4^{+0.4}_{-0.3}$ & $1.8^{+0.8}_{-0.8}$ & $6.3^{+2.1}_{-2.2}$ & $0.6^{+0.4}_{-0.6}$ & $74.5^{+11.0}_{-9.9}$ & $15.7_{-12.7}^{+74.3}$ & $7.7_{-0.4}^{+0.4}$ & 0.1 & 0.2\\

{\bf Grus 2}
& $1.1^{+0.6}_{-0.6}$ & $1.8^{+1.1}_{-1.7}$ & $-1.7^{+2.0}_{-3.4}$ & $-0.3^{+0.4}_{-0.5}$ & $1.7^{+0.8}_{-0.8}$ & $6.5^{+2.1}_{-2.2}$ & $1.0^{+0.6}_{-0.7}$ & $61.4^{+17.7}_{-18.9}$ & 
$18.2_{-18.0}^{+80.2}$&$5.2_{-5.2}^{+1.9}$ & -0.1 & -0.1 \\ 

Hercules
& $1.1^{+0.6}_{-0.6}$ & $3.0^{+1.0}_{-1.2}$ & $-1.4^{+1.8}_{-1.6}$ & $0.2^{+0.6}_{-0.5}$ & $1.8^{+0.8}_{-0.8}$ & $6.3^{+2.1}_{-2.3}$ & $1.0^{+0.7}_{-0.6}$ & $81.8^{+5.7}_{-5.1}$ & 
$114_{-47}^{+86}$& $7.9_{-0.5}^{+0.3}$ & 0.0 & 0.0 \\ 

Horologium I
& $1.1^{+0.6}_{-0.6}$ & $2.2^{+1.1}_{-1.7}$ & $0.2^{+2.3}_{-2.4}$ & $-0.3^{+0.5}_{-0.5}$ & $1.7^{+0.8}_{-0.8}$ & $6.4^{+2.1}_{-2.2}$ & $1.1^{+0.7}_{-0.6}$ & $63.3^{+14.5}_{-17.3}$ &
$558_{-331}^{+1131}$& $7.8_{-2.7}^{+1.1}$ & 0.1 & 0.1 \\ 

Hydra II
& $1.1^{+0.6}_{-0.6}$ & $1.4^{+1.0}_{-1.7}$ & $-2.5^{+1.6}_{-2.8}$ & $-0.4^{+0.4}_{-0.4}$ & $1.7^{+0.8}_{-0.9}$ & $6.6^{+2.2}_{-2.1}$ & $0.8^{+0.6}_{-0.7}$ & $60.0^{+20.1}_{-20.0}$ & 
$0.1_{-0.1}^{+2.8}$& $2.5_{-6.1}^{+3.7}$ & 0.0 & -0.1 \\ 

Leo IV
& $1.0^{+0.6}_{-0.6}$ & $3.3^{+1.3}_{-1.1}$ & $-2.3^{+1.3}_{-2.0}$ & $0.5^{+0.3}_{-0.3}$ & $1.8^{+0.8}_{-0.8}$ & $6.2^{+2.1}_{-2.3}$ & $0.9^{+0.6}_{-0.7}$ & $76.7^{+9.3}_{-8.5}$ &
$16.1_{-11.7}^{+37.5}$& $7.2_{-0.6}^{+0.5}$ & 0.0 & -0.1 \\ 

Leo V
& $1.1^{+0.6}_{-0.6}$ & $3.8^{+0.9}_{-0.8}$ & $-2.3^{+1.2}_{-1.0}$ & $0.2^{+0.7}_{-0.5}$ & $1.9^{+0.8}_{-0.7}$ & $6.3^{+2.1}_{-2.2}$ & $0.6^{+0.4}_{-0.6}$ & $77.0^{+9.2}_{-8.4}$ & 
$6.2_{-5.0}^{+20.6}$& $7.6_{-0.6}^{+0.5}$ & 0.1 & 0.0 \\ 

Leo T
& $1.1^{+0.6}_{-0.6}$ & $3.2^{+1.0}_{-1.1}$ & $-1.4^{+1.6}_{-1.6}$ & $-0.1^{+0.4}_{-0.3}$ & $1.8^{+0.8}_{-0.8}$ & $6.2^{+2.1}_{-2.3}$ & $1.0^{+0.7}_{-0.6}$ & $51.1^{+14.9}_{-22.9}$ & 
$41.9_{-21.6}^{+37.2}$& $8.3_{-0.3}^{+0.3}$ & -0.1 & 0.0 \\ 

Pisces II
& $1.1^{+0.6}_{-0.6}$ & $2.0^{+0.9}_{-1.7}$ & $-0.2^{+2.5}_{-2.9}$ & $-0.2^{+0.5}_{-0.5}$ & $1.7^{+0.8}_{-0.8}$ & $6.4^{+2.1}_{-2.2}$ & $1.1^{+0.7}_{-0.6}$ & $68.5^{+11.7}_{-13.8}$ & 
$209_{-166}^{+767}$& $7.0_{-2.1}^{+1.0}$ & 0.0 & -0.1 \\ 

{\bf Reticulum II}
& $1.1^{+0.6}_{-0.6}$ & $3.3^{+1.2}_{-1.1}$ & $-1.4^{+1.7}_{-1.1}$ & $0.2^{+0.6}_{-0.5}$ & $1.8^{+0.8}_{-0.8}$ & $6.3^{+2.1}_{-2.2}$ & $0.8^{+0.5}_{-0.6}$ & $78.9^{+7.5}_{-7.0}$ & 
$330_{-141}^{+252}$& $8.0_{-0.8}^{+0.5}$ & 0.0 & 0.0\\ 

{\bf Segue 1}
& $1.0^{+0.6}_{-0.6}$ & $2.5^{+1.1}_{-1.4}$ & $-0.9^{+2.4}_{-2.5}$ & $0.3^{+0.3}_{-0.3}$ & $1.6^{+0.8}_{-0.9}$ & $6.2^{+2.1}_{-2.3}$ & $1.6^{+0.6}_{-0.3}$ & $70.6^{+8.2}_{-12.0}$ & 
$2064_{-855}^{+1528}$&$7.9_{-1.6}^{+0.7}$ & 0.0 & 0.0\\ 

{\bf Segue 2}
& $1.2^{+0.6}_{-0.5}$ & $3.1^{+1.3}_{-1.2}$ & $-1.4^{+1.7}_{-1.2}$ & $-0.4^{+0.4}_{-0.5}$ & $1.8^{+0.8}_{-0.8}$ & $6.4^{+2.1}_{-2.2}$ & $0.7^{+0.5}_{-0.7}$ & $61.5^{+21.7}_{-19.4}$ & 
$131_{-119}^{+246}$&$7.7_{-2.4}^{+0.7}$ & 0.1 & 0.2\\ 

Triangulum II
& $1.1^{+0.6}_{-0.6}$ & $1.8^{+1.3}_{-1.9}$ & $-2.4^{+1.7}_{-2.3}$ & $-0.3^{+0.5}_{-0.5}$ & $1.7^{+0.8}_{-0.9}$ & $6.5^{+2.2}_{-2.1}$ & $0.8^{+0.6}_{-0.7}$ & $60.1^{+16.3}_{-19.6}$ & 
$0.3_{-0.3}^{+14.8}$&$4.1_{-6.8}^{+3.0}$ & 0.1 & 0.0 \\ 

Tucana 2
& $1.1^{+0.6}_{-0.6}$ & $2.7^{+0.6}_{-1.1}$ & $-1.3^{+1.8}_{-1.6}$ & $0.0^{+0.4}_{-0.3}$ & $1.7^{+0.8}_{-0.9}$ & $6.3^{+2.1}_{-2.3}$ & $1.0^{+0.7}_{-0.6}$ & $70.9^{+11.1}_{-12.2}$ & $1856_{-846}^{+1787}$ & $7.8_{-0.3}^{+0.4}$ & -0.2 & 0.1 \\

{\bf Tucana 3}
& $1.1^{+0.6}_{-0.6}$ & $1.8^{+1.2}_{-1.8}$ & $-2.4^{+1.7}_{-3.0}$ & $-0.3^{+0.4}_{-0.4}$ & $1.7^{+0.8}_{-0.8}$ & $6.5^{+2.2}_{-2.1}$ & $0.9^{+0.6}_{-0.7}$ & $58.6^{+19.4}_{-20.3}$ & 
$4.5_{-4.5}^{+59.6}$&$4.6_{-6.4}^{+2.3}$ & 0.0 & 0.0\\ 

Tucana 4
& $1.1^{+0.6}_{-0.6}$ & $2.5^{+0.9}_{-1.4}$ & $-1.0^{+2.1}_{-2.2}$ & $-0.0^{+0.5}_{-0.4}$ & $1.7^{+0.8}_{-0.8}$ & $6.3^{+2.1}_{-2.3}$ & $1.1^{+0.7}_{-0.6}$ & $70.1^{+11.1}_{-12.8}$ & 
$1103_{-652}^{+1593}$&$7.5_{-1.2}^{+0.6}$ & 0.1 & 0.0\\ 

{\bf Ursa Major I}
& $1.1^{+0.6}_{-0.6}$ & $3.3^{+1.1}_{-1.1}$ & $-1.6^{+1.7}_{-1.9}$ & $0.6^{+0.2}_{-0.2}$ & $1.7^{+0.8}_{-0.8}$ & $6.2^{+2.1}_{-2.3}$ & $1.1^{+0.7}_{-0.6}$ & $80.2^{+6.6}_{-6.2}$ &
$725_{-265}^{+395}$&$8.2_{-0.2}^{+0.3}$ & 0.1 & 0.3\\ 

{\bf Ursa Major II}
& $1.2^{+0.6}_{-0.5}$ & $3.6^{+1.1}_{-0.9}$ & $-1.3^{+1.5}_{-1.0}$ & $-0.2^{+0.5}_{-0.7}$ & $1.8^{+0.8}_{-0.8}$ & $6.3^{+2.1}_{-2.2}$ & $0.8^{+0.5}_{-0.6}$ & $79.4^{+6.9}_{-6.7}$& 
$977_{-433}^{+834}$& $8.4_{-0.4}^{+0.4}$ & -0.1 & 0.2\\ 

{\bf Willman 1}
& $1.0^{+0.6}_{-0.6}$ & $3.2^{+1.1}_{-1.1}$ & $-1.5^{+1.9}_{-1.7}$ & $0.3^{+0.5}_{-0.3}$ & $1.8^{+0.9}_{-0.8}$ & $6.4^{+2.1}_{-2.2}$ & $1.2^{+0.6}_{-0.5}$ & $75.0^{+9.6}_{-9.7}$ &
$317_{-160}^{+277}$&$8.1_{-1.1}^{+0.6}$ & 0.1 & -0.1\\ 
	\hline
	\end{tabular}
 \end{center}
	%}
\begin{flushleft}
$^{a}$ The Bayes factor (BF) is defined by $P(M_{Q=1}|D)/P(M_\mathrm{best}|D)$ for subscript ``1'', and $P(M_{Q=q^{\prime}}|D)/P(M_\mathrm{best}|D)$ for subscript ``2'', respectively,
where $P(M_{Q=q^{\prime}}|D)$, $P(M_{Q=1}|D)$, and $P(M_\mathrm{best}|D)$ are the mean posterior probabilities of mass models with spherical dark matter halo ($Q=1$), with non-spherical dark matter halo (the axial ratio is fixed as a projected stellar axial ratio: $Q=q^{\prime}$), and with the best posteriors  respectively. In each model, we keep the stellar distributions are axisymmetric shape, that is, stellar axial ratio are not unity. 
$D$ indicates observational data for each dSph.
To confirm this BF analysis, Akaike's Information Criterion with a correction for small sample sizes (c-AIC) is also introduced. We compute c-AIC$_1$ and c-AIC$_2$ (subscript number is the same as the BF) for all dSphs and find that c-AIC$_1\sim$c-AIC$_2\sim4<6$, which means that c-AIC analysis can follow the results from the BF analysis.
\end{flushleft}
\end{table*}

\section{Posterior PDFs} \label{sec:appB}
\begin{figure*}[h!]
	\begin{minipage}{0.49\hsize}
		\begin{center}
			\includegraphics[width=\columnwidth]{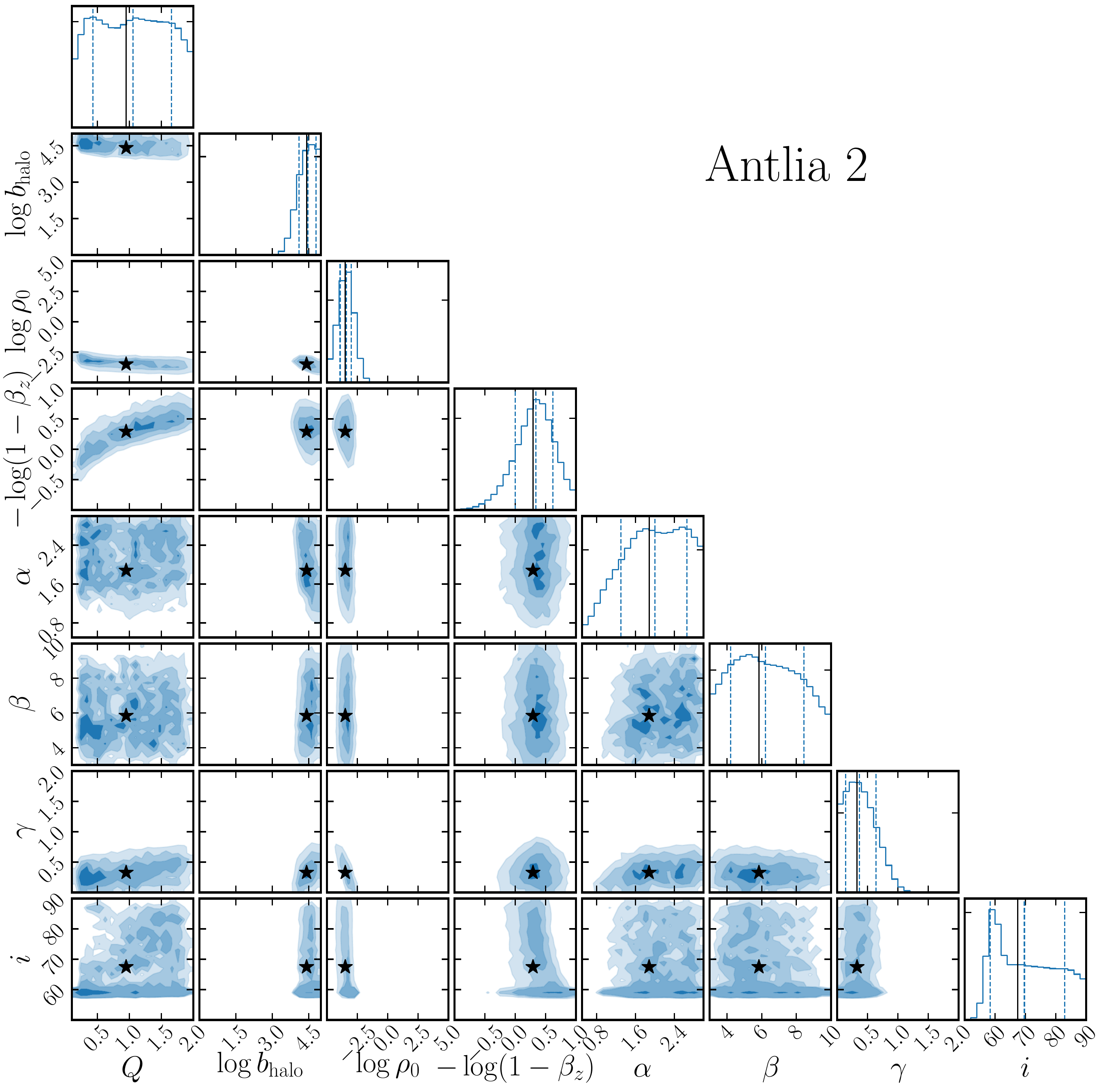}
		\end{center}
	\end{minipage}
	\begin{minipage}{0.49\hsize}
		\begin{center}
			\includegraphics[width=\columnwidth]{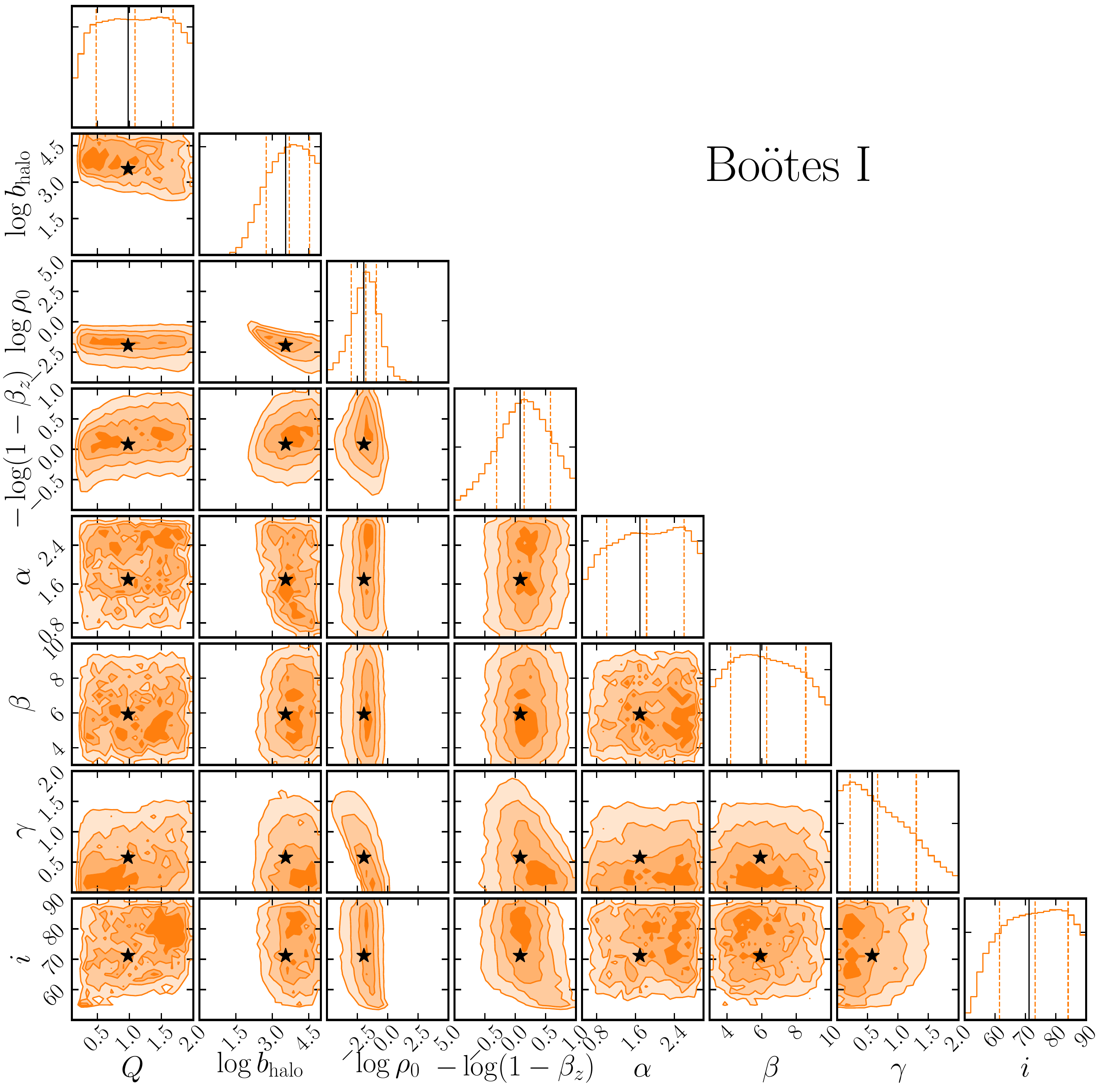}
		\end{center}
	\end{minipage}
	\begin{minipage}{0.49\hsize}
		\begin{center}
			\includegraphics[width=\columnwidth]{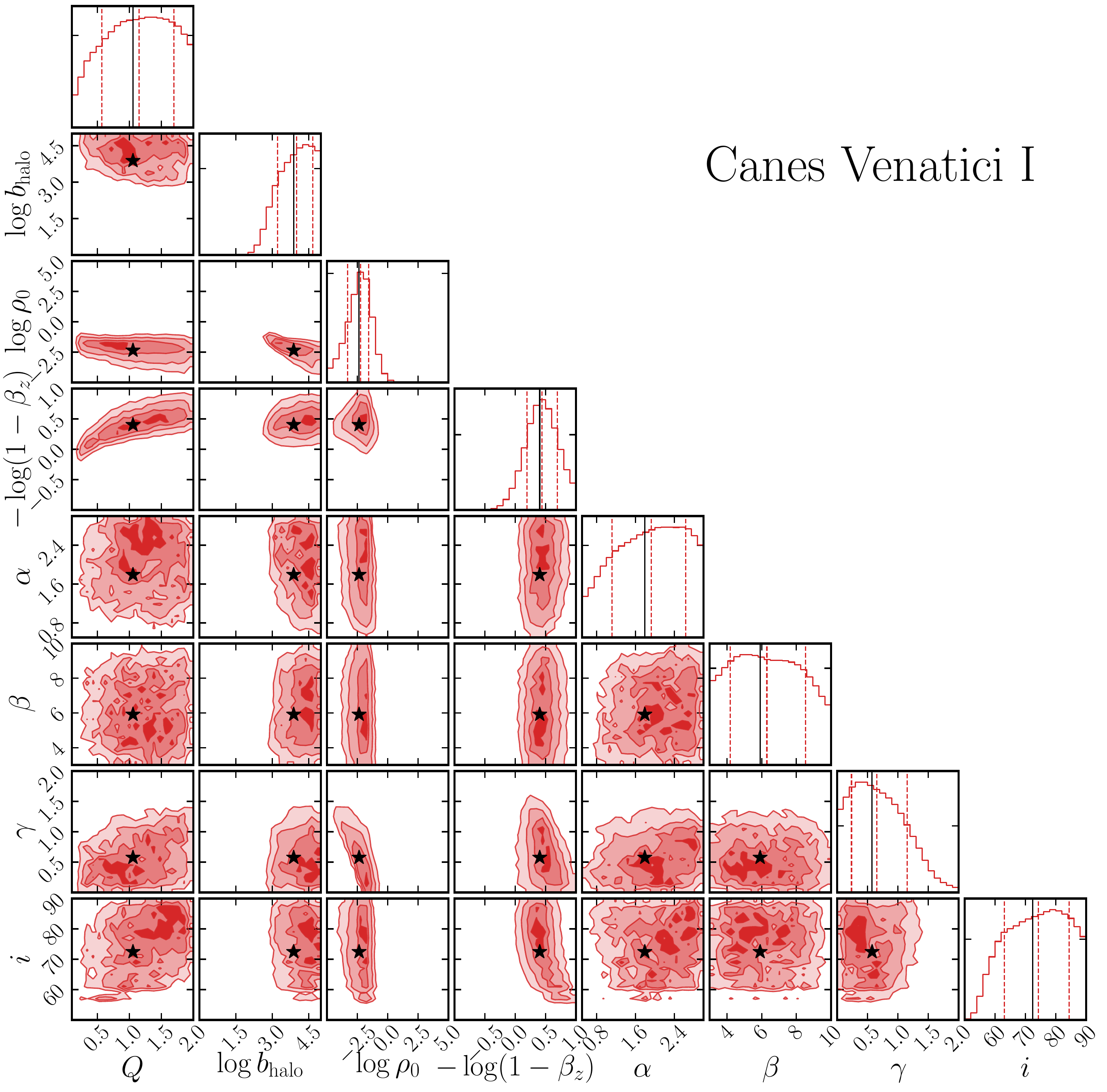}
		\end{center}
	\end{minipage}
	\begin{minipage}{0.49\hsize}
		\begin{center}
			\includegraphics[width=\columnwidth]{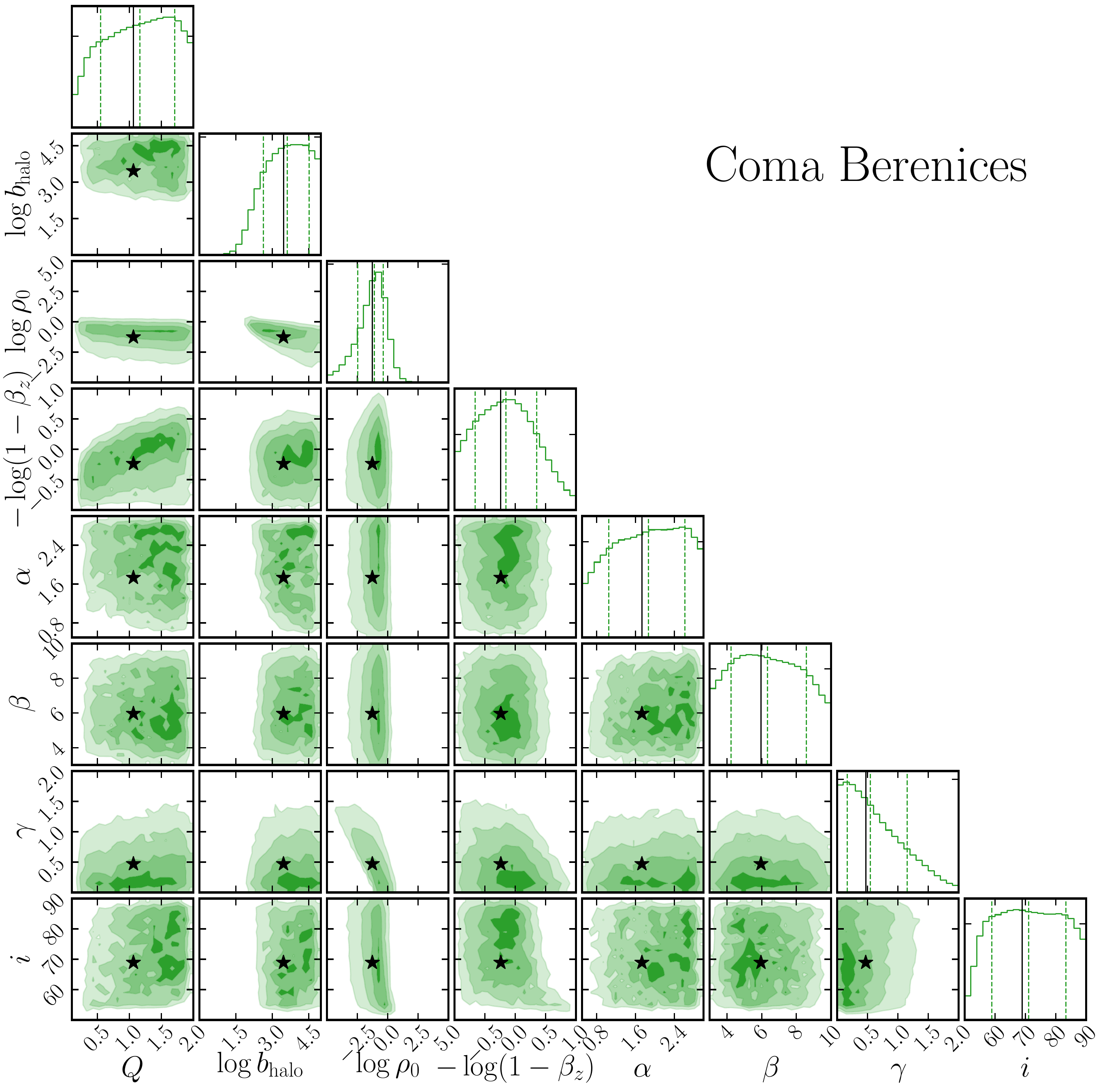}
		\end{center}
	\end{minipage}
    \caption{Posterior probability distributions for the fitting parameters for Antlia~2~(top left), Bo\"otes~I~(top right), Canes~Venatici~I~(bottom left), and Coma~Berenices~(bottom right). The dashed lines in each histogram represent the median and 68~\% confidence values. The contours in each panel are the 68~\%, 95~\%, and 99.7~\% ~(from dark to light) confidence regions. 
    The black stars and vertical solid lines show the parameters with the highest posteriors.}
    Units of $b_\mathrm{halo}$ and $\rho_0$ are [pc] and $[M_\odot$~pc$^{-3}]$, respectively.
    \label{PDFs1}
\end{figure*}

\begin{figure*}[h!]
	\begin{minipage}{0.49\hsize}
		\begin{center}
			\includegraphics[width=\columnwidth]{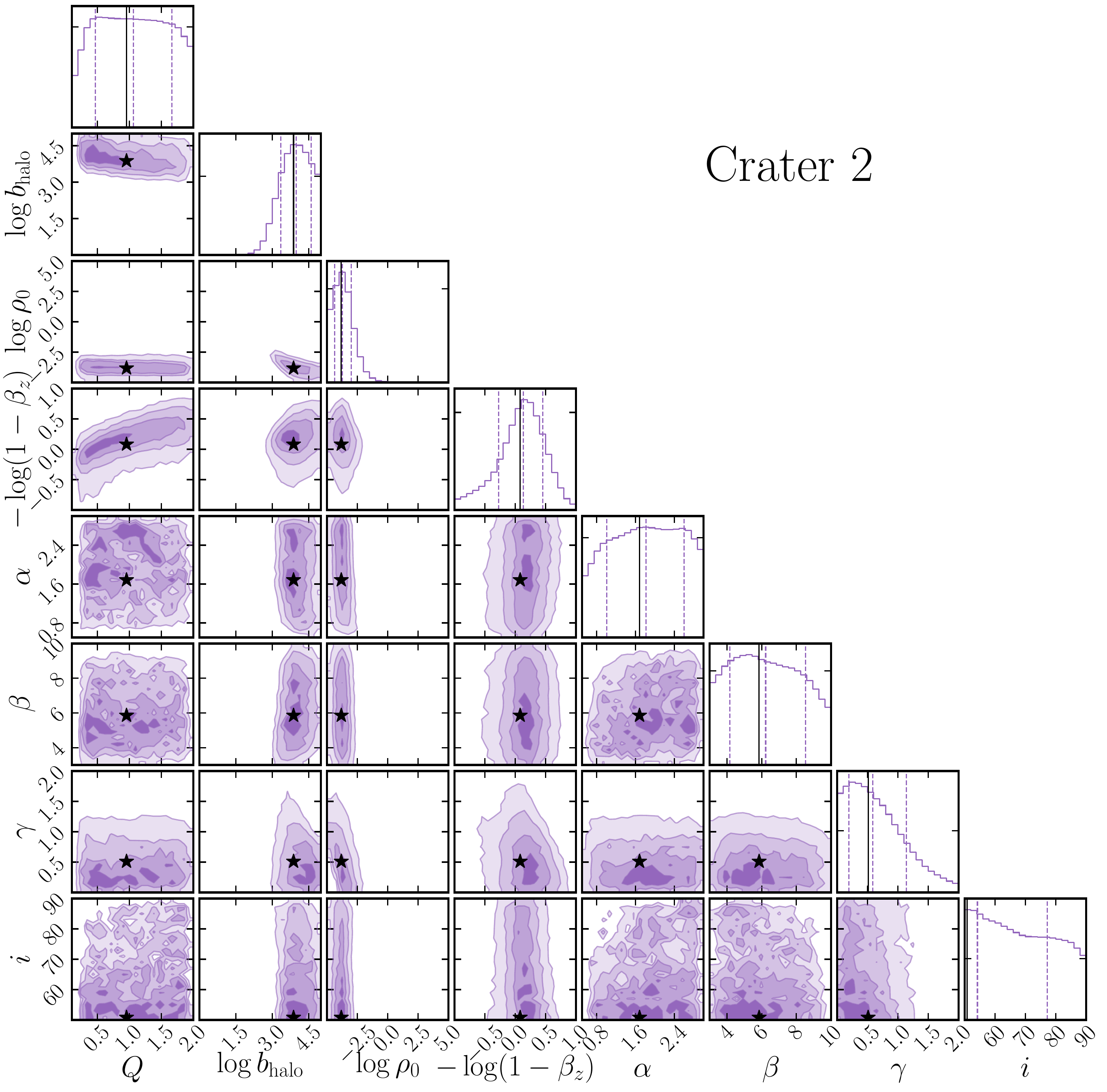}
		\end{center}
	\end{minipage}
	\begin{minipage}{0.49\hsize}
		\begin{center}
			\includegraphics[width=\columnwidth]{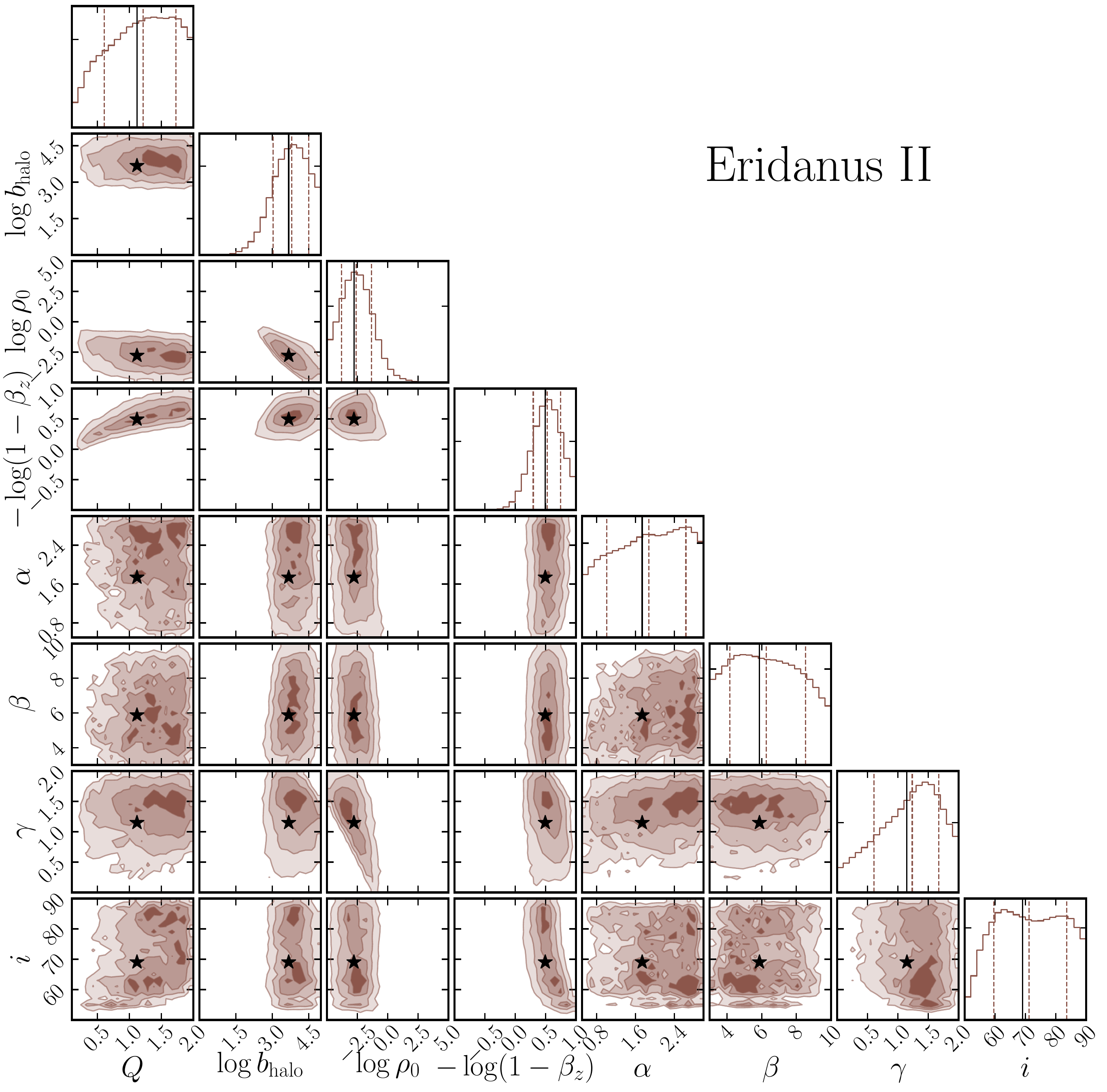}
		\end{center}
	\end{minipage}
	\begin{minipage}{0.49\hsize}
		\begin{center}
			\includegraphics[width=\columnwidth]{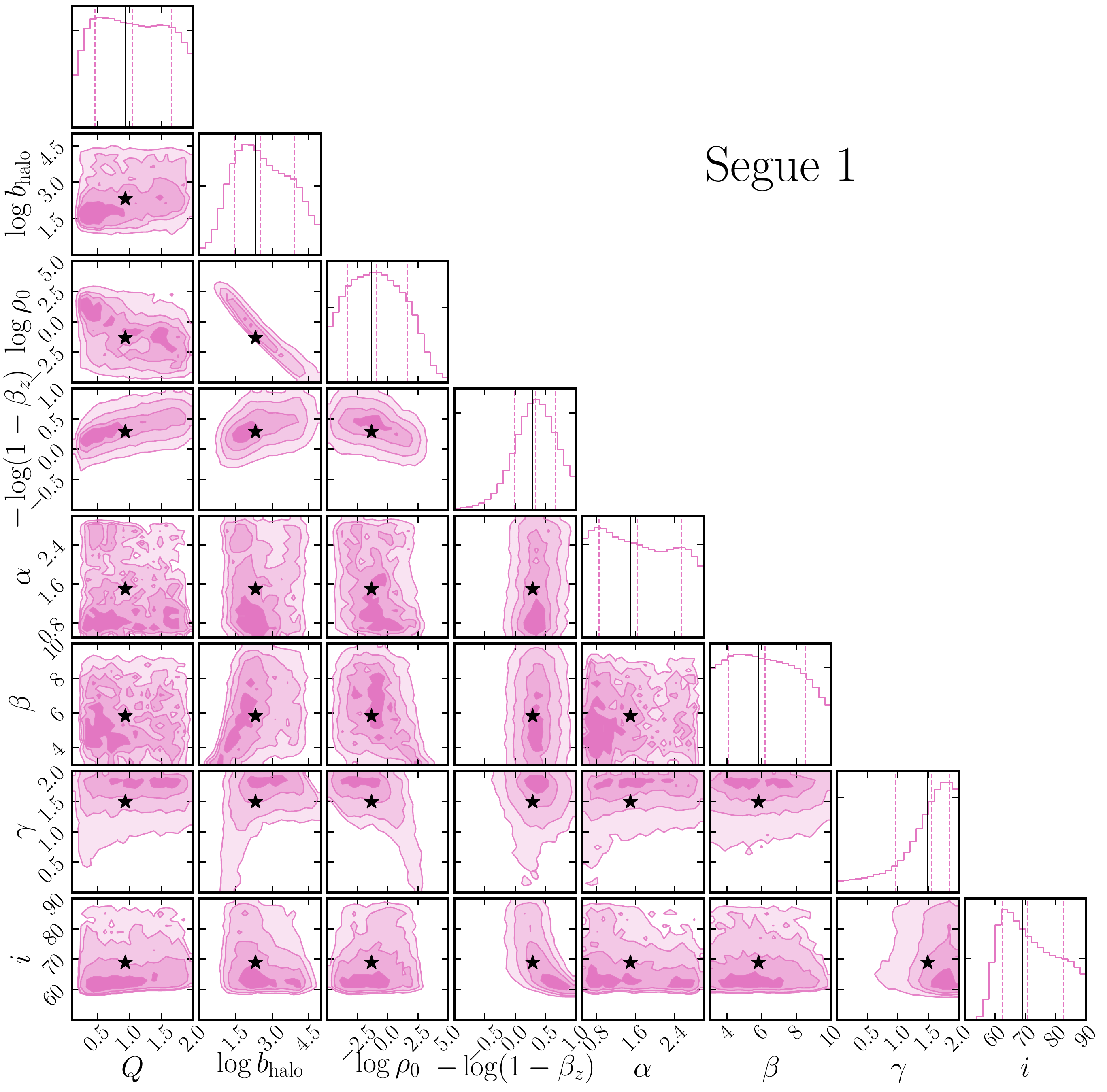}
		\end{center}
	\end{minipage}
	\begin{minipage}{0.49\hsize}
		\begin{center}
			\includegraphics[width=\columnwidth]{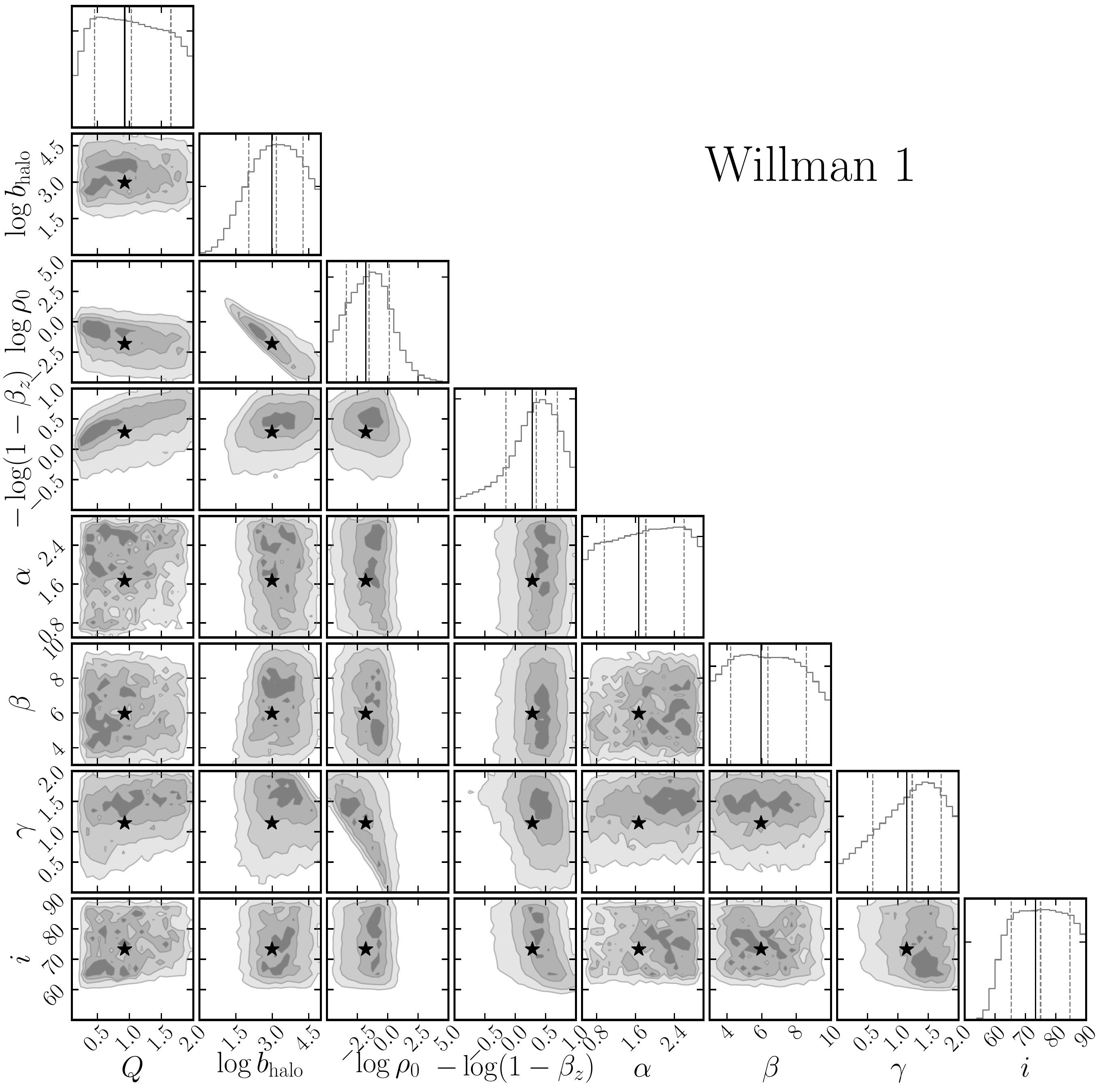}
		\end{center}
	\end{minipage}
    \caption{Same as figure~\ref{PDFs1}, but for Crater~2~(top left), Eridanus~II~(top right), Segue~1 (bottom left), and Willman~1~(bottom right).}
    \label{PDFs2}
\end{figure*}

Here we show the results of parameter estimations based on an MCMC fitting analysis.
Figures~\ref{PDFs1} and \ref{PDFs2} display the posterior PDFs for Antlia~2, Bo\"otes~I, Canes Venatici~I, Coma~Berenices, Crater~2, Eridanus~II, Segue~1, and Willman~1 as the representative satellites, which have the largest kinematics samples among our sample in this work. 

\clearpage
\bibliographystyle{aasjournal}
\bibliography{reference}{}

\end{document}